\documentclass[1p]{Article_website}

\usepackage{amsmath}
\usepackage{amsthm}
\usepackage{amsfonts}
\usepackage{placeins}
\usepackage{multicol}
\usepackage[usenames,dvipsnames,svgnames,table]{xcolor}
\usepackage{epstopdf}

\usepackage{soul}

\newlength\figswidth
\figswidth=\textwidth

\begin{document}

\begin{frontmatter}

\title{Stabilization of acoustic modes using Helmholtz and Quarter-Wave resonators tuned at exceptional points}

\author[myaddress]{C. Bourquard}
\author[myaddress]{N. Noiray\corref{mycorrespondingauthor}}
\cortext[mycorrespondingauthor]{Corresponding author}
\ead{noirayn@ethz.ch}

\address[myaddress]{CAPS Laboratory, Department of Mechanical and Process Engineering, ETH Zurich, Zurich 8092, Switzerland}

\begin{abstract}
Acoustic dampers are efficient and cost-effective means for suppressing thermoacoustic instabilities in combustion chambers. However, their design and the choice of their purging air mass flow is a challenging task, when one aims at ensuring thermoacoustic stability after their implementation. In the present experimental and theoretical study, Helmholtz (HH) and Quarter-Wave (QW) dampers are considered. A model for their acoustic impedance is derived and experimentally validated. In a second part, a thermoacoustic instability is mimicked by an electro-acoustic feedback loop in a rectangular cavity, to which the dampers are added. The length of the dampers can be adjusted, so that the system can be studied for tuned and detuned conditions. The stability of the coupled system is investigated experimentally and then analytically, which shows that for tuned dampers, the best stabilization is achieved at the exceptional point. The stabilization capabilities of HH and QW dampers are compared for given damper volume and purge mass flow.
\end{abstract}


\end{frontmatter}

\begin{table}[!ht]
\begin{center}
\begin{footnotesize}
\begin{tabular}{|p{0.8cm} p{6.5cm} p{0.5cm} p{6.0cm}|}\hline
& & & \\
\multicolumn{2}{|l}{\textbf{\normalsize{Nomenclature}}} & &\\
& & & \\
$a$ 			& cross sectional area of the HH neck  &
$S_D$			& area of the chamber covered with dampers\\
$A$				& HH volume and QW cross sectional area & 
$S_\text{IT}$	& cross sectional area of the impedance tube\\
$A_\eta$		& acoustic amplitude of the limit cycle  &
$St$			& Strouhal number ($St =  \omega_H \sqrt{4a/\pi}/\bar{u}$)\\
$c$				& speed of sound & 
$\bar{u}$		& purge flow velocity\\
$c_1$			& linear gain applied to the voltage signal & 
$u\, (\hat{u})$	& acoustic velocity in time (frequency) domain\\
$c_2$			& cubic saturation coefficient of the voltage signal &
$U$				& voltage signal from the feedback microphone\\
$D$				& characteristic polynomial of the coupled system &
$V$				& volume of the chamber\\
$e_{1,2}$		& eigenvectors of the coupled system &
$V_Q$			& volume of the QW\\
$f$				& frequency & 
$Z$				& wall impedance\\
$f_0$			& frequency of the mode of interest & 
$Z_d$			& damper impedance\\
$f_\text{wd}$	& eigenfrequency of the chamber with dampers &
$Z_{H,Q}$		& impedance of the HH/QW damper\\
$g$				& approximation function for cotangent around $\pi/2$ &
$\alpha$		& linear damping coefficient of the chamber\\
$\mathcal{H}_{\text{wod},\text{wd}}$ & transfer function without and with dampers &
$\alpha_{H,Q}$	& linear damping coefficient of the HH/QW damper\\
$K_{Q}$ 		& acoustic stiffness of the QW &
$\beta$			& linear gain of the electro-acoustic feedback loop\\
$K_\text{ac}$ 	& acoustic stiffness of air column under bulk compression & 
$\gamma$		& specific heat ratio\\
$l$				& effective HH neck length &
$\delta$		& detuning ($\delta=(\omega_H-\omega_0)/\omega_0$)\\
$l_\text{cor}$	& end-correction on both sides of the HH neck &
$\delta_b$		& acoustic boundary layer thickness ($\delta_b=\sqrt{2\nu/\omega}$)\\
$l_p$			& physical HH neck length &
$\varepsilon_{H,Q}$& damping efficiency factor\\
$\mathcal{L}_\text{bl}$	& acoustic power loss per unit area in the acoustic boundary layer & 
$\zeta_{H,Q}$	& pressure loss coefficient of the HH/QW damper\\
$L$				& QW effective length / length of the HH back cavity &
$\eta$ & amplitude of the dominant acoustic mode, with $p(t,\boldsymbol{x})\approx\eta(t)\psi(\boldsymbol{x})$\\
$L_p$			& QW physical length &
$\kappa$		& effective cubic saturation constant\\
$L_\text{cor}$	& QW end-correction & 
$\lambda$		& eigenvalue of the coupled system\\
$\dot{m}$		& air mass flow to purge the damper& 
$\Lambda$		& norm of the dominant acoustic mode \\
$\dot{m}_{0,H/Q}$ & purge mass flow for which $\mathcal{R}=0$ & 
$\nu$			& kinematic viscosity\\
$\bar{p}$		& ambient pressure & 
$\nu_\text{wod}$& linear growth/decay rate of the dominant acoustic mode in the chamber \textit{without} dampers\\
$p\, (\hat{p})$	& acoustic pressure in time (frequency) domain & 
$\nu_\text{wd}$	& linear growth/decay rate of the dominant acoustic mode in the chamber \textit{with} dampers\\
$\mathcal{P}_L$	& total power loss in the acoustic boundary layer & 
$\rho$ 			& air density\\
$Pr$			& Prandtl number & 
$\sigma_{H,Q}$  & porosity ($\sigma_H = a/S_\text{IT}$ and $\sigma_Q=A/S_\text{IT}$)\\
$\hat{Q}_{C,N}$ & coherent/noisy component of the acoustic source in the chamber volume& 
$\tau$			& time delay of the electroacoustic feedback\\
$r$				& QW radius & 
$\psi$			& shape of the dominant acoustic mode\\
$R_{bl}$		& boundary layer resistance of the QW & 
$\Psi_d$		& weighting coefficient (position of damper w.r.t. the mode shape)\\
$R_{H,Q}$		& resistance of the HH/QW damper & 
$\omega$		& angular frequency\\
$R_{vs}$		& vortex shedding resistance of the QW & 
$\omega_0$		& natural eigenfrequency of the dominant acoustic mode\\
$\mathcal{R}$	& reflection coefficient & 
$\omega_{H,Q}$	& eigenfrequency of the HH/QW damper\\
$s$				& Laplace variable & 
$\omega_\text{wd}$	& eigenfrequency of the coupled system (chamber with damper)\\
& & & \\
\hline
\end{tabular}
\end{footnotesize}
\end{center}
\end{table}

\section{Introduction}
\noindent In order to achieve efficient and clean combustion, there is a dire need for robust control strategies to prevent thermoacoustic instabilities. The use of passive damping devices such as Helmholtz (HH) or Quarter-Wave (QW) dampers is a cost-effective option to prevent these combustion instabilities \cite{zhao2015pas,lahiri2017jsv}. 
In the seventies, thermoacoustic instabilities in rocket engines were the topic of several studies, where different acoustic damping enhancement strategies were compared: for example baffles, HH and cylindrical liners in \cite{crocco1969cs}, HH, QW and Quincke resonator in \cite{harrje1972nasa}. A more recent study dealing with the comparison of absorption coefficients of half-wave, QW and HH is proposed in \cite{sohn2011ast}. Keller \cite{keller1974nasa} stated that the QW has a narrower bandwidth than the HH, but it is less influenced by nonlinearities \cite{laudien1995paa}. Although up until two decades ago, the studies on damping device design for rocket engine combustion instabilities still dealt with QW as well as HH \cite{acker1994jpp}, nowadays most of the available literature concentrates on QW resonator rings \cite{oschwald2008jpp}: in \cite{oschwald2011ppp}, the influence of the resonator length on the frequency of the engine acoustic modes is studied and \cite{cardenas2014jpp} presents a graphical method based on a low-order network model to determine the stability of the system. The influence of such a QW resonator ring on the shape of the longitudinal and transversal \cite{schulze2015jsr} as well as azimuthal \cite{zahn2016asme} modes, and on the stability margin of the engine \cite{zahn2017asme} has also been studied. On the contrary, the literature on QW applications in aeroengines is sparse (e.g. \cite{joshi1998asme} and \cite{mongia2003jpp} with QW tubes installed upstream of the premixers), whereas one can find many papers about acoustic liners, behaving as matrices of HH resonators \cite{garrison1971propconf,hughes1990jfm,burak2009aiaa}. HH and QW are also used to hinder thermoacoustic instabilities in the combustors of land-based gas turbines for power generation \cite{richards2003jpp}. Over the last two decades most publications deal with the use of HH dampers, either conventional \cite{schlein1999asme} (with a detailed model derived in \cite{bellucci2004jegtp} and design principles given in \cite{dupere2005jegtp}), or featuring multiple volumes, i.e. having  several interconnected inner cavities  \cite{bothien2013jegtp,bothien2013asme}. Overall, the choice of using HH or QW dampers to suppress thermoaocustic instabilities in combustion chambers is often guided by field-specific trends or  past experience of the manufacturers. The present study provides a detailed comparison of the damping capabilities of HH and QW resonators.\\
Another aspect of the present work is to investigate the stabilizing capabilities of flow-purged dampers on self-sustained oscillations. A large number of investigations dealing with modal damping enhancement using acoustic dampers generally focus on linearly stable configurations where there is no through-flow in the damper: for instance, in \cite{esteve2002jasa,pietrzko2008ast}, the optimal damping is determined for a configuration where acoustic and structural modes are coupled; In \cite{yu2008jasa,klaus2014aa}, the authors underline that the optimal damping depends on whether one wants to achieve minimum narrow-band or broad-band response or minimum reverberation time. The influence of the detuning of the damper has been studied in \cite{cossalter1994icec,gysling2000asme} while the effect of multiple dampers is scrutinized in \cite{soon2012ksnve}. Several experimental and analytical studies deal with methods to find the optimum number of dampers and their best positioning in the combustion chamber, e.g. \cite{yu2009jsv,zalluhoglu2017jdsmc}. A recent work proposes to automatize the damper design process by using computationally-cheap adjoint-based optimization \cite{mensah2017jegtp}.\\
In the specific case of gas turbine combustors, dampers and perforated liners are connected to the combustion chamber. Their neck interfaces the hot combustion products, and the dampers are usually air-purged (e.g. \cite{dupere2005jegtp,zhong2012jasa}), in order to adjust their acoustic resistance, and to prevent hot gas ingestion, which could not only damage them, but also detune them. A few studies deal with the influence of the associated density discontinuity on the impedance of Helmholtz dampers \cite{bothien2015aiaa,yang2017aiaa}. Another recent work \cite{cosic2012jegtp} investigates how the impedance of a HH damper nonlinearly depends on the amplitude of the acoustic level in the combustion chamber, which induces, beyond a certain threshold, \emph{periodic} hot gas ingestion.\\
In this context, the goal of the present experimental and analytical study is to build on the work proposed in \cite{noiray2012jsv} and investigate the potential of air-purged HH and QW dampers to increase modal damping in combustion chambers. This investigation focuses on the linear stability of the coupled system ``dampers-combustion chamber'', and the reader can refer to \cite{bourquard2018asme} for the complementary study  dealing with the associated nonlinear dynamics.
In the first part, the impedance of stand-alone HH and QW dampers is modeled and experimentally validated for a range of purge mass flows and damper volumes. In practice, the available ranges for these two parameters are bounded by technical constraints: i) the available volume for damper implementation is usually limited and the size of the dampers must be adjusted accordingly; ii) the amount of purge air must be as low as possible, but sufficiently large to provide required damping performance. In fact, regarding ii), the following conflicting constraints must be satisfied: bypassing compressed air from the combustion process to supply the dampers has a negative impact on the performance and emissions of the combustor and it should therefore be minimized; however, it should be large enough such that the risk of hot gas ingestion is properly mitigated.\\
In the second part of the paper, the dampers are connected to a chamber in order to stabilize self-sustained acoustic oscillations. For the experimental investigation, one uses an electro-acoustic feedback in an enclosure in order to mimic thermoacoustic instabilities in combustion chambers. This experimental set-up is more  flexible than a combustion experiment and gives full control on the parameters governing the self-sustained acoustic oscillations. A theoretical model of the coupled system ``chamber-damper'' is derived and successfully compared against experimental data.\\
In the last part of this work, one investigates the existence of acoustic \textit{exceptional points} in chambers which are equipped with dissipative resonators. Exceptional points (EP) pertain to systems exhibiting a special eigenvalue degeneracy, for which not only the eigenvalues, but also the eigenvectors coalesce when one of the governing parameters is adjusted. Investigation of EPs in quantum mechanics, optics, electronics, mechanics or acoustics is the subject of intense ongoing research \cite{shi2016nc}, e.g.  \cite{hoffmann2003zamm} for damping of friction-induced instabilities, 
\cite{zhu2014prx} for unidirectional invisibility in an acoustic waveguide,  \cite{gao2015n} for  exciton-polaritons in  semiconductor microcavities, \cite{brandstetter2014nc} for coupled lasers,   \cite{achilleos2017prb,xiong2017jasa} for the design of acoustic metamaterials,  \cite{ryu2015pre} for the transient dynamics in the vicinity of EPs, or \cite{ding2016prx} for the intriguing acoustical properties of coupled cavities. Their importance in the understanding of thermoacoustic instabilities has been highlighted in \cite{mensah2018jsv}.
In the present study, it will be shown that the best stabilization of the acoustic mode is achieved  when the resonance frequency and the damping of the HH or QW resonators are fine-tuned at the EP of the coupled system.

\section{Damper modelling}

\subsection{Impedance model}
\noindent In this section second order harmonic oscillator models are introduced for the impedance of HH and QW dampers. In the remainder of the article, $(\cdot)_H$ refers to quantities related to the HH resonator, while $(\cdot)_Q$ refers to those related to the QW resonator. Both resonators considered in the present study are axisymmetric.
\begin{figure}
\centering
\includegraphics[width=\figswidth]{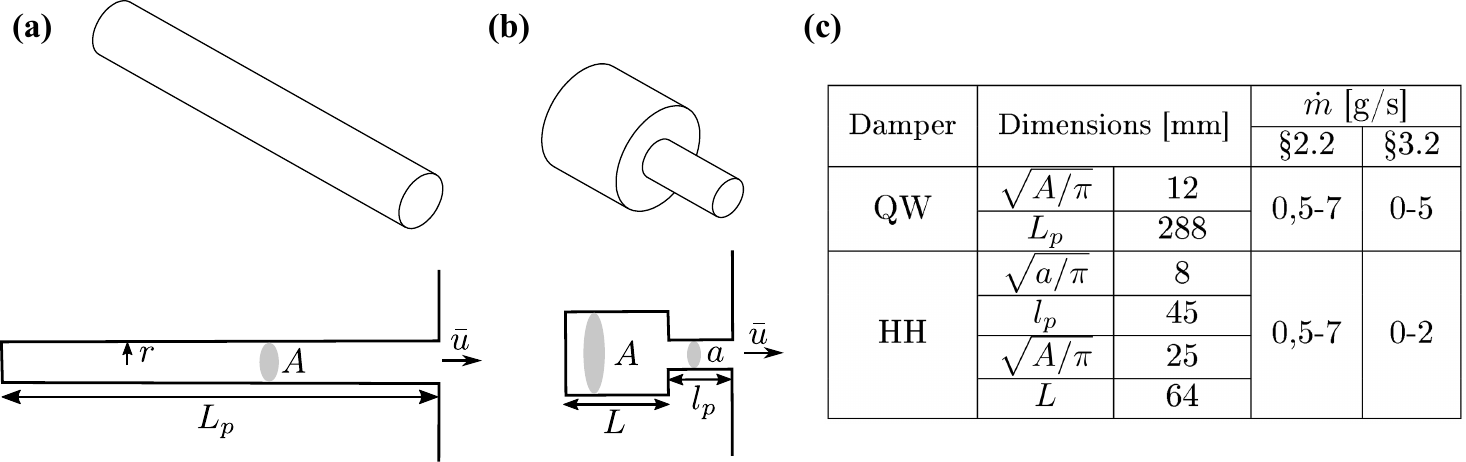}
\caption{Resonator geometry for \textbf{(a)} QW, \textbf{(b)} HH. \textbf{(c)} Table summarizing the dimensions of the dampers and the mass flows used throughout the paper.}
\label{fig:dampers}
\end{figure}
\subsubsection{Helmholtz resonator (H)}
\noindent The HH resonator is sketched in Fig. \ref{fig:dampers}b and Fig. \ref{fig:imptube}b. Assuming plane wave propagation in the back volume of length $L$ and in the neck of effective length $l=l_p + l_\text{cor}$ (with end corrections on both sides), one can write the following expression for the damper reactance at the neck:

\begin{equation}
\Im (Z_H) = -\rho c \,\,\frac{a - A\tan(\omega l/c)\tan(\omega L/c)}{a\tan(\omega l/c)+ A\tan(\omega L/c)},
\label{eq:tot_imp}
\end{equation}

\noindent with $Z_H$ the damper impedance, $\rho$ and $c$ the air density and the speed of sound in the damper, $\omega$ the angular frequency, $a$ the neck cross-section and $A$ the back-volume cross-section. Considering compact neck and volume length with respect to the resonance wavelength   ($\tan{\omega L/c} \simeq \omega L/c$ and $\tan{\omega l/c} \simeq \omega l/c$), and the fact that the area ratio between neck and volume is small ($a\ll A$), Eq. \ref{eq:tot_imp} can be simplified to:

\begin{equation}
\Im (Z_H) = \rho l \, \frac{\omega^2-\omega_{H}^2}{\omega},
\label{eq:impH_s_nores}
\end{equation}

\noindent where $\omega_{H} = c \sqrt{a / A L l}$ the damper's resonance frequency. 
In the present work, it is assumed that coherent vortex shedding at the HH resonator mouth is the main dissipation mechanism \cite{tam2001jsv} and that the resistance of the damper can be written as:

\begin{equation}
\Re(Z_H)=R_{H}=\zeta_H \rho \bar{u} = \zeta_H \frac{\dot{m}}{a},
\label{eq:res_H}
\end{equation}

\noindent with  $\zeta_H$ a pressure loss coefficient depending on the  neck geometry and position, $\bar{u}$ the mean velocity through the HH neck, and $\dot{m}$ the mean mass flow through the neck, which is a critical parameter in real turbomachinery applications. This expression is obtained by linearizing the Bernoulli equation across the neck (e.g. \cite{rienstra2015book}). Please note that should the acoustic amplitude become high, the resistance would not be proportional to $\bar{u}$ but to $|\bar{u}+u'|$, with $u'$ the acoustic velocity in the neck. Using a purely linear dissipation term is justified for the present study, which focuses on the linear stability limits. The reader can refer to \cite{bourquard2018asme} where the nonlinear problem is investigated. Eq. \ref{eq:res_H} does not depend on the resonance frequency, and  the pressure loss coefficient $\zeta_H$ gives the energy transfer from the acoustically-driven incompressible potential flow through the neck to the vortices that are periodically shed from the rim of the neck outlet. One can refer to \cite{yang2016jsv} for a detailed investigation on the modeling of orifice impedance for a broad range of geometries and Strouhal numbers. Combining \eqref{eq:impH_s_nores} and \eqref{eq:res_H}, one obtains the HH resonator impedance:

\begin{equation}
Z_{H} = \rho l\, \frac{s^2 + \omega_{H}^2}{s} +R_H,
\label{eq:impH_s_res}
\end{equation}

\noindent with $s = i\omega$ the Laplace variable, which is the classical ($l-\zeta$) model, where $l$ stands for the effective length of the inertial mass of air in the orifice and $\zeta$ for the pressure loss coefficient defining the acoustic resistance against the  oscillation of air in the orifice. It was discussed for example by Morse and Ingard \cite{morse1968book} at page 760, or in \cite{bellucci2004jegtp,rienstra2015book}.

\subsubsection{Quarter-wave resonator (Q)}
\noindent The QW resonator is sketched in Fig. \ref{fig:dampers}a and Fig. \ref{fig:imptube}c. Assuming plane wave propagation in the resonator with an effective length $L=L_p + L_\text{cor}$ (including an end correction at the outlet), one gets the following expression for the damper reactance: 

\begin{equation}
\Im(Z_{Q}) = - \rho c\,\,\frac{1}{\tan(\omega L/c)}.
\label{eq:imp_QW}
\end{equation}

\noindent At resonance frequency, $\omega_Q L/c = \pi/2$. For angular frequencies $\omega$ that are close to the resonance frequency, one can use the following approximation:

\begin{equation}
\dfrac{1}{\tan x} \simeq g(x) = -\frac{1}{2} \left( x - \frac{\pi^2}{4x}\right),
\label{eq:simp_func_QW}
\end{equation}

\begin{figure}[t]
\centering
\includegraphics[width=0.4\figswidth]{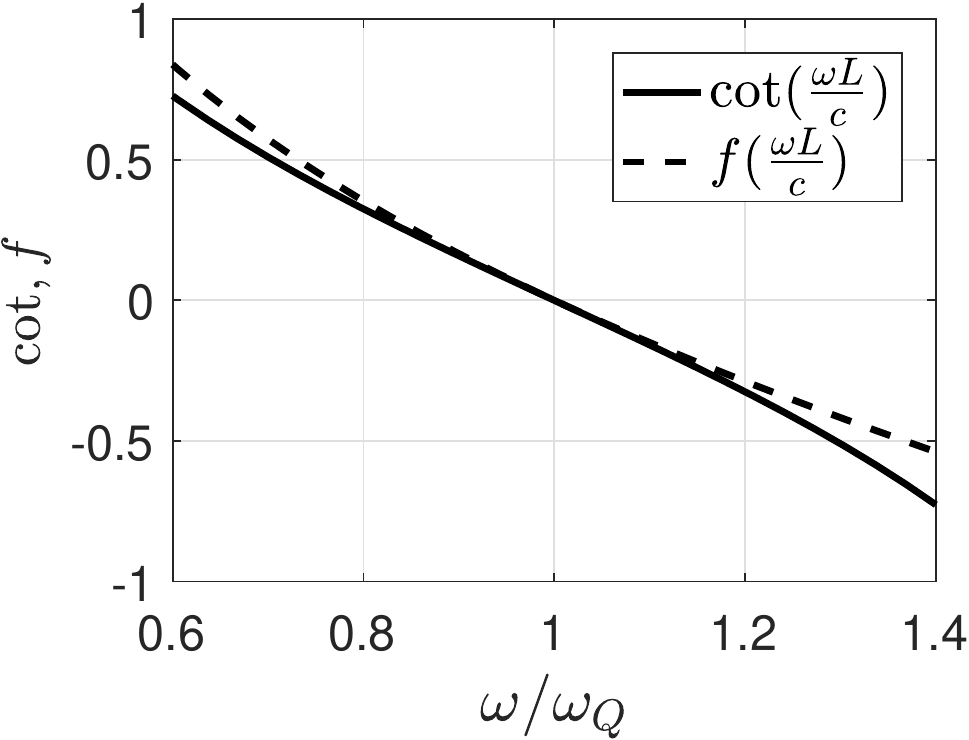}
\caption{Comparison between actual reactance with cotangent and simplified function. $\omega_Q=\pi c/2L$ so that when $\omega=\omega_Q$, $\omega L/c=\pi/2$}
\label{fig:cotan}
\end{figure}

\noindent which is illustrated in Fig. \ref{fig:cotan} and leads to:

\begin{equation}
\Im(Z_{Q}) =\rho \frac{L}{2} \, \frac{\omega^2 - \omega_{Q}^2}{\omega},
\label{eq:QW_impedance}
\end{equation}

\noindent with  $\omega_{Q} = \pi c /2 L$ the QW resonance frequency. Using this approximation is quite uncommon in the literature, although it provides an explicit  formulation of the QW resonator impedance as a second order harmonic oscillator. Regarding dissipation, the resistive term in the QW case is composed of two contributions: one of them is the vortex shedding at the damper mouth as for the HH resonator:

\begin{equation}
R_{vs}=\rho \zeta_Q \bar{u} = \zeta_Q \frac{\dot{m}}{A}.
\end{equation} 

\noindent The second contribution comes from the losses in the acoustic boundary layer \cite{morse1968book,searby2008jpp}, with the following expression for the viscous and thermal power loss per wall unit area:

\begin{equation}
\mathcal{L}_\text{bl}= \dfrac{\rho\, \omega\, \delta_\nu}{2} \, |\hat{u}_\text{rms}|^2 + \dfrac{(\gamma-1)\,\rho\,\omega \,\delta_\nu}{2\sqrt{Pr}}  \, \left|\dfrac{\hat{p}_\text{rms}}{\rho c}\right|^2,
\label{eq:P_loss}
\end{equation}

\noindent where $\hat{u}_\text{rms}$ and $\hat{p}_\text{rms}$ are the root mean square amplitude of the acoustic velocity and pressure in the tube, $\delta_\nu=\sqrt{2 \nu/\omega}$ the acoustic boundary layer thickness with $\nu$ the kinematic viscosity equal to $1.5 \cdot 10^{-5}$ m\textsuperscript{2}/s in air at ambient condition. $\gamma$ is the specific heat ratio, with $\gamma \bar{p}=\rho c^2$, and $Pr$ is the Prandtl number equal to $0.71$ for air. Using the acoustic velocity and pressure distribution along the tube:

\begin{equation}
\hat{u}_\text{rms} (x) = \dfrac{|\hat{u}_\text{max}|}{\sqrt{2}} \sin \left(\dfrac{x \pi}{2 L} \right) \quad \text{and} \quad \hat{p}_\text{rms} (x) = \dfrac{|\hat{p}_\text{max}|}{\sqrt{2}} \cos \left(\dfrac{x \pi}{2 L} \right),
\end{equation}

\noindent using $|\hat{p}_\text{max}|=\rho c |\hat{u}_\text{max}|$, multiplying by the perimeter $2\pi r$ and integrating over the physical length $L_p$ with respect to $x$ gives the total boundary layer power losses for the QW damper:

\begin{equation}
\mathcal{P}_L = \rho \dfrac{\pi r L_p}{4} \omega \delta_\nu \left(1+\dfrac{\gamma-1}{\sqrt{Pr}}\right) |\hat{u}_\text{max}|^2.
\label{eq:P_loss_whole_QW}
\end{equation}

\noindent Note that the viscous losses are about twice as high as the thermal losses. Hence, in the absence of mean flow, the acoustic resistance per unit area  associated with the above power loss is: 

\begin{equation}
R_{bl} = \frac{\mathcal{P}_L}{\pi r^2}\frac{2}{|\hat{u}_\text{max}|^2}=\,\rho \dfrac{L_p}{2 r} \sqrt{2\nu\omega}\, \left(1+\dfrac{\gamma-1}{\sqrt{Pr}}\right).
\label{eq:R_bl}
\end{equation}

\noindent Since $R_{bl}$ is proportional to $\sqrt{\omega}$ one approximates a constant value around the resonance frequency of the damper. The total resistive term is then:

\begin{equation}
R_Q=R_{bl} + R_{vs}= \rho \frac{L_p}{2r} \sqrt{2\nu\omega_Q} \, \left(1+\dfrac{\gamma-1}{\sqrt{Pr}}\right) + \zeta_Q \frac{\dot{m}}{A},
\label{eq:res_Q}
\end{equation}

\noindent and the QW damper impedance:

\begin{equation}
Z_{Q} =\rho \frac{L}{2} \cdot \frac{s^2 + \omega_{Q}^2}{s} + R_Q.
\label{eq:QW_impedance_res}
\end{equation}

\noindent Using the impedance equations and accounting for the interface area ($a$ for HH, $A$ for QW), one can get the resistance, mass and stiffness of the equivalent mechanical oscillators given in Table \ref{tab:equivalent_imp}.

\begin{table}[t]
\begin{center}
\def\arraystretch{2}
\begin{tabular}{|c|c|c|c|}
\hline
& Resistance {\footnotesize [kg.$s^{-1}$]} & Mass {\footnotesize [kg]} & Stiffness {\footnotesize [kg.$s^{-2}$]}\\
\hline
QW & $ \rho \dfrac{AL_p}{2r} \sqrt{2\nu\omega_Q} \, \left(1+\dfrac{\gamma-1}{\sqrt{Pr}}\right)  + \zeta_Q \dot{m}$ & $ \dfrac{\rho L A}{2}$ & $\dfrac{\pi^2}{8}\dfrac{\rho c^2 A}{L}$ \\[1.2ex]
\hline
HH & $\zeta_H \dot{m}$ & $\rho l a$ & $\dfrac{\rho c^2 a^2}{A L}$ \\[1.2ex]
\hline
\end{tabular}
\caption{Equivalent resistance, mass and stiffness of HH and QW resonators.}
\label{tab:equivalent_imp}
\end{center}
\end{table}

\noindent Note that the equivalent stiffness of the QW $K_{Q}$ is slightly higher than the one $K_\text{ac}$ associated with the bulk compression of an air column of length $L$:

\begin{equation}
K_{Q} = \dfrac{\pi^2}{8}\dfrac{\rho c^2 A}{L} = \dfrac{\pi^2}{8}\dfrac{\gamma \bar{p} A}{L} = \dfrac{\pi^2}{8} K_\text{ac},
\end{equation}

\noindent with $\bar{p}$ is the ambient pressure.

\subsection{Reflection coefficient measurements and model tuning}
\label{subsec:R_meas}
\begin{figure}[h]
\centering
\includegraphics[width=\figswidth]{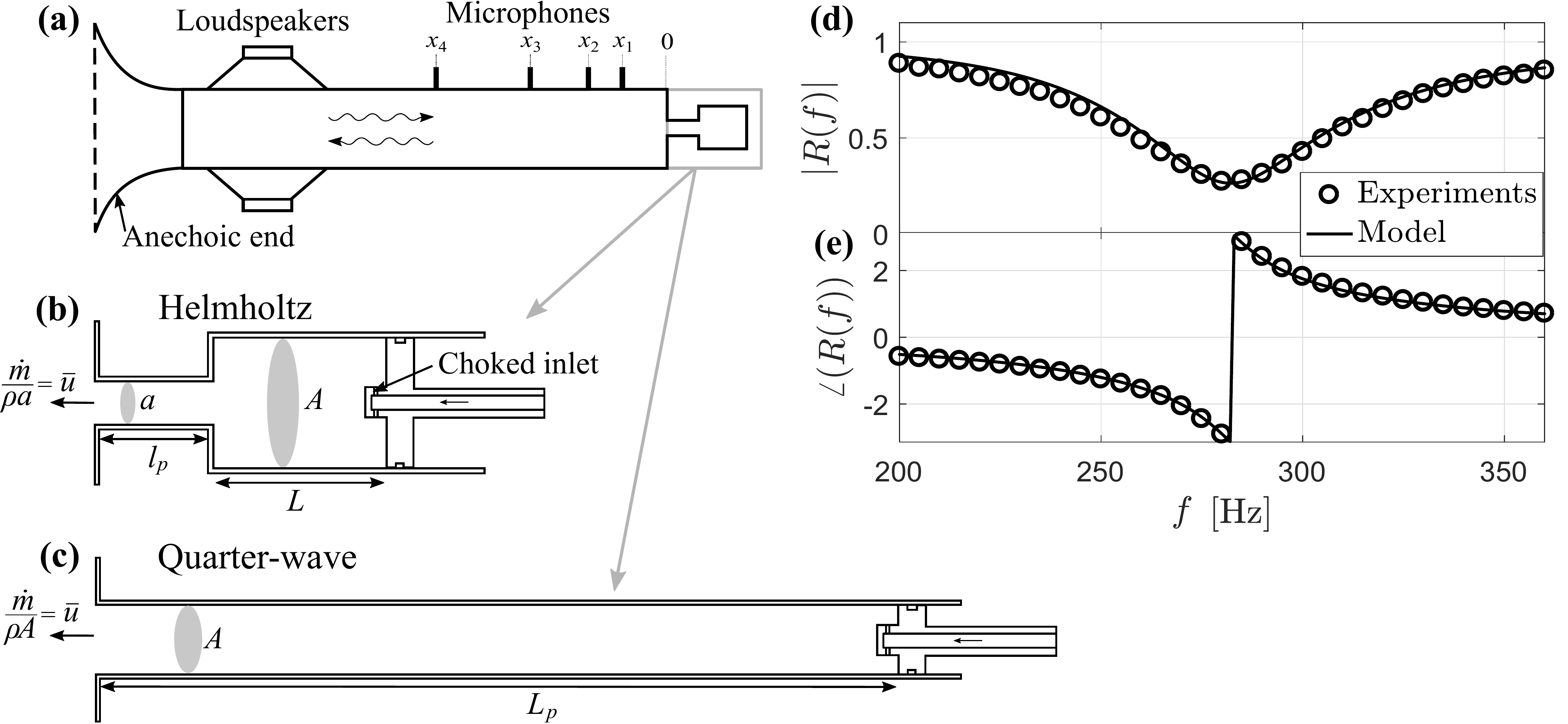}
\caption{\textbf{(a)} Schematic of the impedance tube used for the reflection coefficient measurements. \textbf{(b)} Sketch of the Helmholtz resonator with variable back volume length. \textbf{(c)} Sketch of the Quarter-Wave resonator with variable length. \textbf{(d)} Magnitude and \textbf{(e)} phase of the reflection coefficient for the HH damper with $\dot{m}=1.5$ g/s. The model (plain line) is fitted to the experimental results (circles) by adjusting $\zeta_H$.}
\label{fig:imptube}
\end{figure}

\noindent A HH damper and a QW damper were used for the experimental investigations. A piston allows variation of their volume such that their resonance frequency can be adjusted between 200 and 500 Hz. The HH damper neck has a diameter of 16 mm and a length of 45 mm. At the frequency of interest in the second part of the paper (287 Hz), the back-volume of the HH is 64 mm long, with a diameter of 50 mm (see Fig. \ref{fig:imptube}b), and the length of the QW is 288 mm with a diameter of 24 mm (see Fig. \ref{fig:imptube}c). For the reflection coefficients measurements, the mass flow $\dot{m}$ is varied between 0.5 and 7 g/s.

\begin{figure}[t]
\centering
\includegraphics[width=\figswidth]{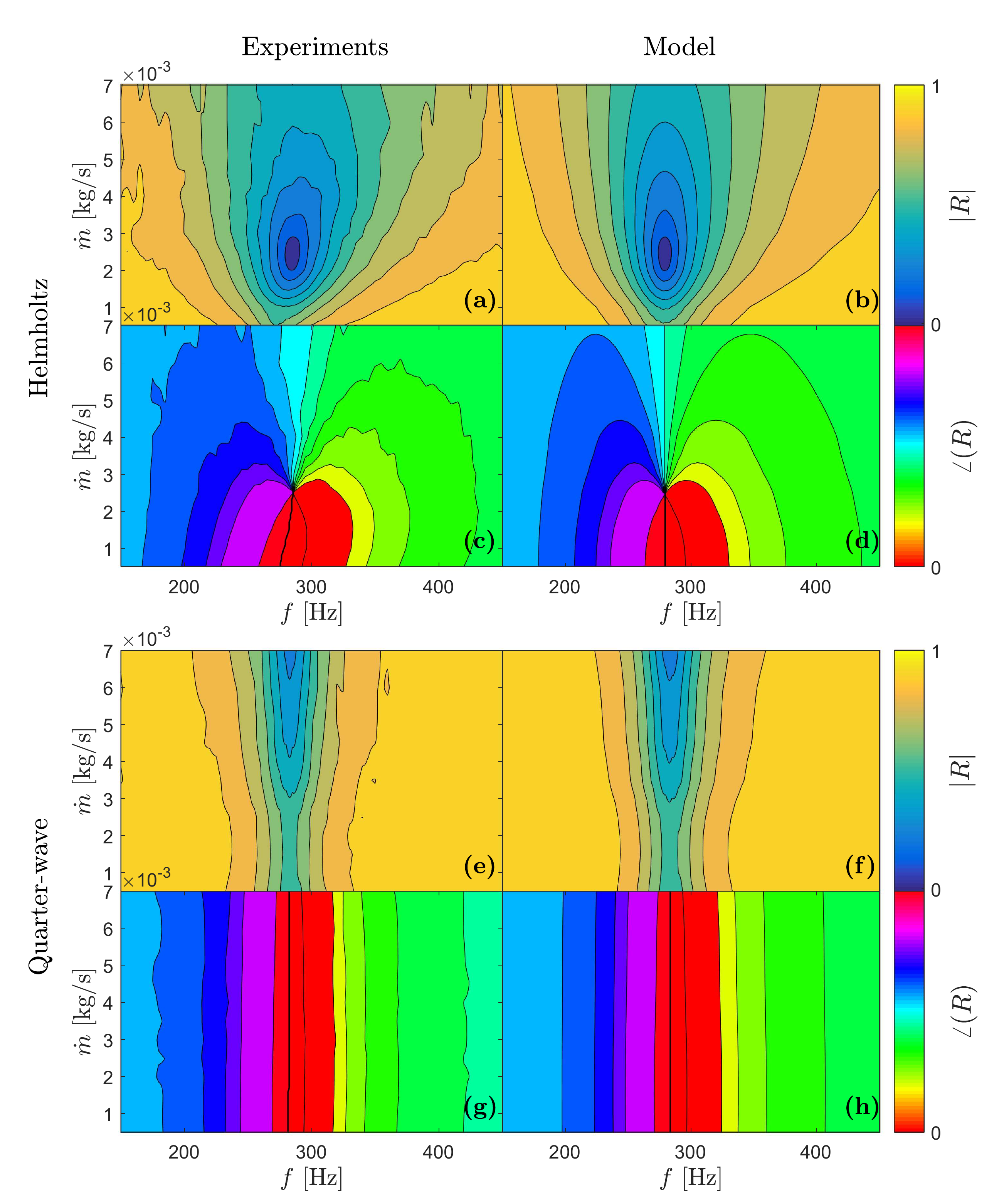}
\caption{Results of the reflection coefficient measurements (left) and the tuned model (right). For the Helmholtz resonator: magnitude (\textbf{(a)}, \textbf{(b)}) and phase (\textbf{(c)}, \textbf{(d)}). For the Quarter-Wave resonator: magnitude (\textbf{(e)}, \textbf{(f)}) and phase (\textbf{(g)}, \textbf{(h)}).}
\label{fig:R_coeff}
\end{figure}

\noindent The impedances of the HH and QW dampers given in Eqs \ref{eq:impH_s_res} and \ref{eq:QW_impedance_res} feature two parameters that are empirically estimated in the present work: the length corrections ($l_\text{cor}$ and $L_\text{cor}$ respectively) and the pressure loss terms ($\zeta_H$ and $\zeta_Q$ respectively). \\
The end corrections are determined using the Helmholtz solver AVSP for a configuration where the dampers are connected to an impedance tube with a 62$\times$62 mm$^2$ cross-section. AVSP solves the Helmholtz equation as an eigenvalue problem in quiescent domains with possible non-uniform temperature distribution \cite{nicoud2007aiaa}. The solver gives the first eigenfrequency and the end corrections are obtained using the expressions $\omega_{H} = c \sqrt{a / A L l}$ and $\omega_{Q} = \pi c /2 L$. This gives $l_{\text{cor}}=13.2$ mm and $L_{\text{cor}}=4.2$ mm.\\
For the determination of the pressure loss coefficients $\zeta_H$ and $\zeta_Q$, reflection coefficients measurements were performed with an impedance tube of section $S_\text{IT}=$62$\times$62 mm\textsuperscript{2} (see Fig. \ref{fig:imptube}a) using the Multi-Microphone-Method (MMM) \cite{schuermans2004asme}. For each mass flow, the value of the acoustic resistance is empirically adjusted so that the best fit between experimental and  theoretical reflection coefficient $\mathcal{R}=(Z - \rho c)/(Z + \rho c)$ is achieved. Here, the analytical expression for the impedance $Z$ is obtained by dividing Eq. \ref{eq:impH_s_res} for the HH (resp. Eq. \ref{eq:QW_impedance_res} for the QW) by the area ratio $\sigma_H = a/S_\text{IT}$ (resp. $\sigma_Q=A/S_\text{IT}$), such that it can be quantitatively compared to the experiments. An example of comparison between the HH damper model with tuned parameter $R_H$ and the measurements for a selected mass flow $\dot{m}=1.5$ g/s is given in Fig. \ref{fig:imptube}d and \ref{fig:imptube}e. Figure \ref{fig:R_coeff}a-d shows the comparison between the model and the experiments for a range of mass flows. The comparison is also made for the QW damper in Fig. \ref{fig:R_coeff}e-h. Overall, there is good agreement between model and experiments. One can note that there is a small drift of the eigenfrequency of the HH damper as a function of the purge mass flow, which means that the latter has an influence on the end correction, and that the HH damper is more prone to detuning than the QW damper when the velocity in the neck changes.

\begin{figure}[t]
\centering
\includegraphics[width=0.4\figswidth]{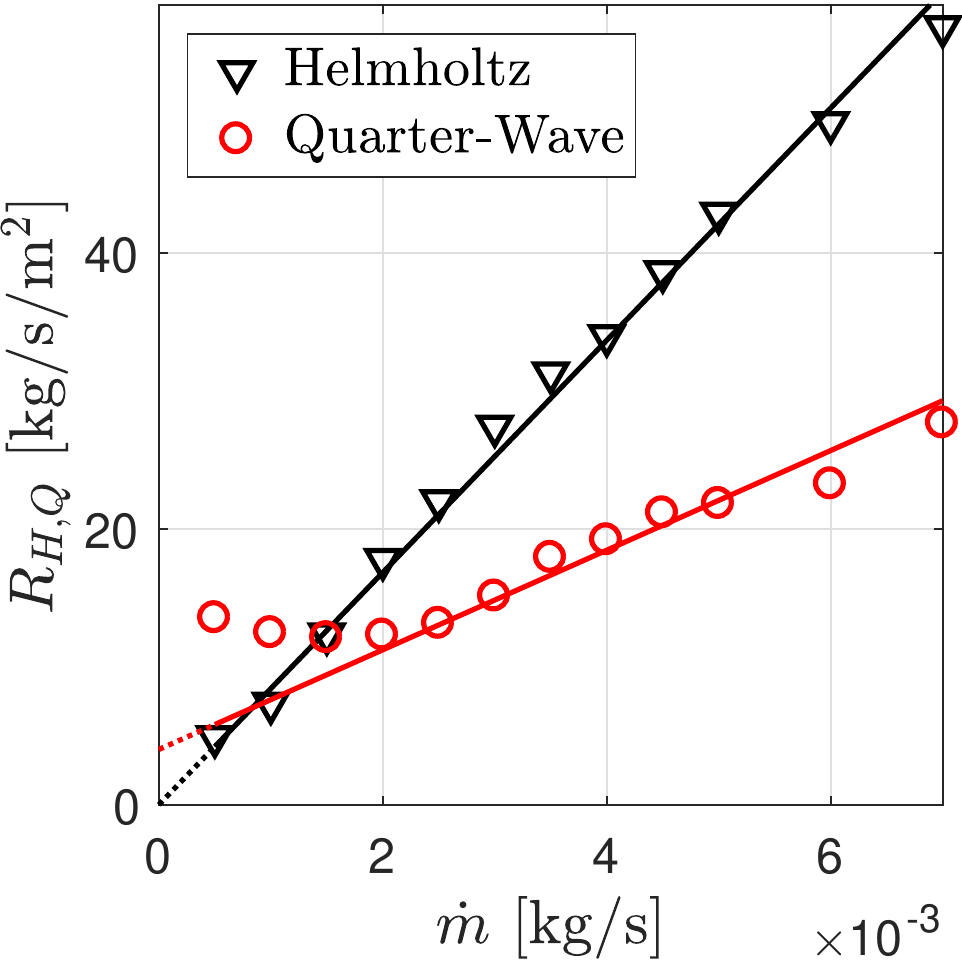}
\caption{Acoustic resistance deduced from the measurement of the reflection coefficient as function of mass flow for the HH (triangles) and the QW (circles). Plain lines correspond to the linear regression on the value of $\zeta_H$ and $\zeta_Q$ using the analytical expressions from Eqs. \ref{eq:res_H} and \ref{eq:res_Q}: $R_{H}= \zeta_H \frac{\dot{m}}{a}$ and $R_Q=\rho \frac{L}{2r} \sqrt{\omega_Q} \, \theta_b + \zeta_Q \frac{\dot{m}}{A}$. For the QW damper, the acoustic boundary layer losses do not depend on the mass flow, which is why the red line does not cross the origin.}
\label{fig:zetafit}
\end{figure}

\noindent A linear regression is then used in both cases to determine the optimum values of $\zeta_H$ and $\zeta_Q$ from the fitted resistive terms. This is shown in Fig. \ref{fig:zetafit} and one obtains $\zeta_H=1.78$ and $\zeta_Q=1.64$. \\
The mass flow $\dot{m}_0$ for which the reflection coefficient vanishes, which corresponds to an anechoic condition in the impedance tube, can be easily determined by matching the impedance at the plane where the damper is connected to the characteristic impedance of air. This mass flow depends on the cross-section of the impedance tube used for the measurement and is not an intrinsic property of the damper. At the resonance frequency, the reactive part of $Z_H$ and $Z_Q$ is zero. The condition giving no reflection is then: $R_H / \sigma_H =\rho c$ and $R_Q / \sigma_Q =\rho c$. Using $c=343$ m/s and $\rho=1.14$ kg.m\textsuperscript{-3} (the measurements were done at 500m above sea level) gives $\dot{m}_{0,H}=2.6$ g/s which is in good agreement with the experiments,  and $\dot{m}_{0,Q}=11.3$ g/s.\\
For the HH damper this is also consistent with the findings of Scarpato et al. \cite{scarpato2012jsv} who state that, at the anechoic condition, the Mach number divided by the porosity increases monotonically from 0.5 for low Strouhal number to $2/\pi$ for high Strouhal number. In the present work, the Strouhal number based on the opening diameter ($St = \omega_H  2\sqrt{a/\pi}/\bar{u}$) ranges from 3 to 15, which correspond to a ``high Strouhal'' regime. Therefore, applying the condition proposed in \cite{scarpato2012jsv}:

\begin{equation}
\frac{\bar{u}}{c} \frac{1}{\sigma_H}= \frac{\dot{m}_{0,H}}{\rho\, a\, c} \frac{1}{\sigma_H} = \frac{2}{\pi}\\
\label{eq:bestresterm_R}
\end{equation}

\noindent gives $\dot{m}_{0,H} = 2.9 $ g/s, which is in good agreement with the experimental value.

\FloatBarrier
\section{Coupled damper-cavity experiments}
\label{sec:experiments}
\begin{figure}[t]
\centering
\includegraphics[width=\figswidth]{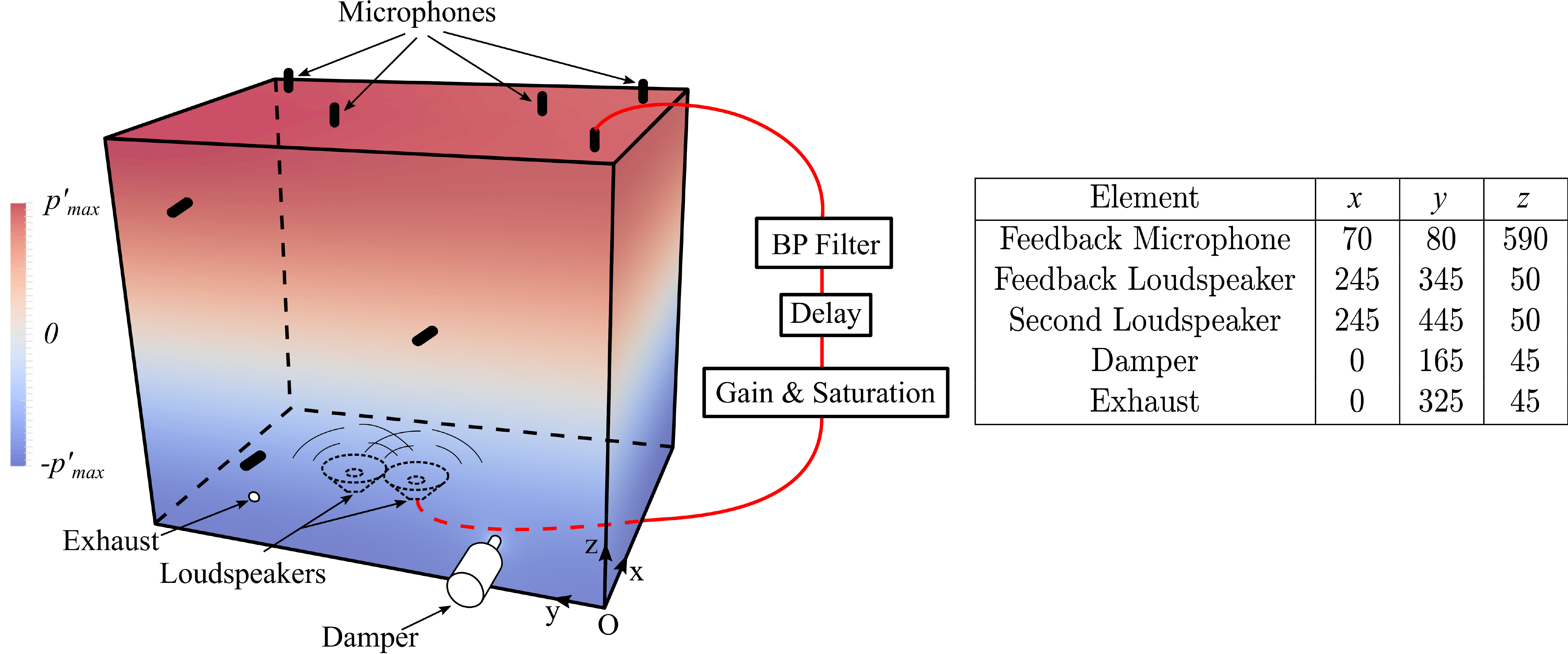}
\caption{Sketch of the experimental setup and coordinates of the elements in mm}
\label{fig:cavity_setup}
\end{figure}

\noindent The problem of dampers that are connected to a combustion chamber is now considered. This is a classical acoustic problem of coupled cavities (e.g. \cite{morse1968book}, Chap. 10.4), in the specific situation where one of the cavities, the combustion chamber, has a volume that is much larger than the one of the secondary cavities, namely the dampers. In that particular situation, the shapes of the first eigenmodes in the main cavity are usually not significantly altered by the implementation of the dampers, but the latter can strongly impact the stability of these modes.  \\ 
In the present work, an electroacoustic feedback in a 0.2 m$^3$ chamber is used to mimic the thermoacoustic coupling occurring in combustion chambers. This experimental set-up allows for a precise control of the linear stability of one of the eigenmodes, with and without dampers. It is sketched in Fig. \ref{fig:cavity_setup}.

\subsection{Stand-alone cavity characterization}
\noindent The experimental setup is composed of: 1) a rectangular metal box (500$\times$700$\times$600 mm$^3$) with stiffening ribs on the outer side of the walls to prevent strong vibro-acoustic feedback; 2) a set of eight G.R.A.S. 46BD 1/4" CCP microphones distributed on 2 of the faces of the cavity; 3) two Pioneer TS-1001I loudspeakers placed inside the cavity; 4) an electro-acoustic feedback loop, which consists in band-pass filtering, delaying and amplifying the signal from one of the microphones and delivering the  output signal to one of the loudspeakers. By varying the amplification and the delay in the feedback loop, it is possible to vary the linear stability of one of the acoustic modes of the main cavity. The second loudspeaker serves as external excitation for forced experiments. The coordinates of the different elements are given in Fig. \ref{fig:cavity_setup}. The microphones are calibrated using the Norsonic Nor1251 calibrator, giving an output of 114dB at 1000Hz.

\begin{figure}[t]
\centering
\includegraphics[width=\figswidth]{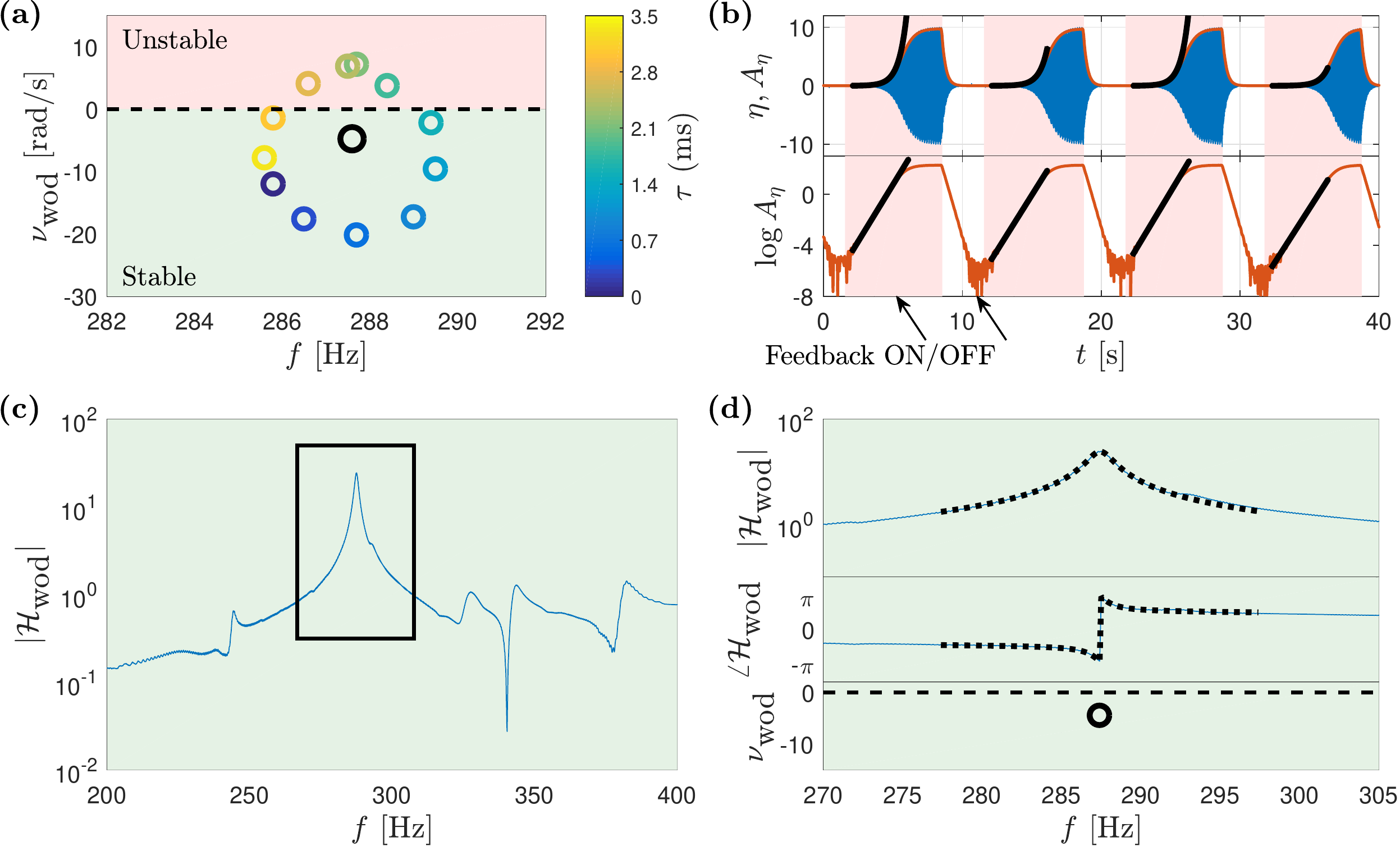}
\caption{\textbf{(a)} Measured eigenfrequencies of the first transversal mode, without and with electroacoustic feedback (resp. black and colored circles), for a range of feedback loop delays, and for the cavity without dampers. \textbf{(b)} For feedback loop delays $\tau$ leading to linearly unstable situations, the linear growth rate (real part of the eigenfrequency) are obtained by fitting an exponential on the transient. \textbf{(c)} For feedback loop delays $\tau$ leading to linearly stable situations, the second loudspeaker is used to force the system with a frequency sweep. The associated transfer function $\mathcal{H}_\text{wod}$ shown in \textbf{(c)} and \textbf{(d)} is fitted with a second-order low-pass filter, which provides the eigenfrequency of the electro-acoustic system.}
\label{fig:meas_tech}
\end{figure}

\noindent The acoustic eigenmode considered in the following sections of the paper is the first transversal mode (see Fig.~\ref{fig:cavity_setup}). The natural damping $\alpha$ of this mode, without damper and without electro-acoustic feedback, is obtained by imposing a ten-second linear sweep excitation from $200$ to $400$Hz using the second loudspeaker. A 2\textsuperscript{nd} order transfer function is then fitted to the experimental transfer function between one of the microphone signals and the excitation signal in order to determine the eigenfrequency and the corresponding damping rate. This procedure is illustrated in Fig. \ref{fig:meas_tech}c and \ref{fig:meas_tech}d and it provides the natural damping $\alpha=-4.8$ rad/s, which is also indicated as the central black dot in Fig. \ref{fig:meas_tech}a. The main contribution to the damping of the eigenmode is attributed in the present situation to one-way acoustic-structure interaction. In the following part of this work, a QW or HH damper is mounted  on the cavity and a small hole in one of the walls of the cavity serves as exhaust for the damper purge air that is injected into the cavity (see Fig. \ref{fig:cavity_setup}). Damping rate measurements were performed in order to identify the effect of an air flow through this exhaust orifice on the natural damping of this eigenmode. These tests showed that the corresponding additional damping does not exceeds 0.4 rad/s ($\alpha$ comprised between -4.8 and -5.2 rad/s) over the entire range of purge mass flows considered, which is negligible compared to the damping rate variations induced by the electro-acoustic feedback loop and/or the HH and QW dampers.

\noindent The signal of the microphone used in the feedback loop is filtered, delayed, amplified to supply the loudspeaker as it was done in e.g. \cite{noiray2012jsv,noiray2013ijnlm}. This real-time signal processing is done using  a NI cRIO-9066 board. The filter is necessary to ensure that the feedback loop only acts on the first transversal acoustic mode of the cavity  at a frequency of about $287$ Hz. The filter is a bandpass with cut-off frequencies at 260 and 320 Hz. When the gain and the time delay in the feedback loop lead to a linearly unstable situation, there is an exponential growth of the amplitude of the first transversal mode. A nonlinear cubic term is implemented in the feedback loop such that the exponentially growing oscillations saturate on a limit cycle, with the dynamics of a Van der Pol oscillator. One can therefore express the  amplitude  $\eta$ of the first transversal mode of the electro-acoustic system as:

\begin{equation}
\ddot{\eta} - (2\nu_\text{wod} + \kappa  \eta^2) \dot{\eta} + \omega_0^2\eta = 0,
\label{eq:VdP}
\end{equation}

\noindent with $\nu_\text{wod}$ its linear growth/decay rate without damper, $\omega_0$ its natural angular frequency and $\kappa$ the saturation coefficient. The linear growth rate $\nu_\text{wod}$ results from the contribution of the linear damping $\alpha$ and the linear gain of the electro-acoustic feedback loop $\beta$, such that $\nu_\text{wod}=\beta-\alpha$. The input of the loudspeaker is proportional to $c_1 U(t+\tau) + c_2 U^3(t+\tau)$, where $U$ is the output voltage of the band-pass filtered microphone signal, $c_1$ and $c_2$ are the linear amplification and the saturation coefficients, and $\tau$ the feedback delay. The latter three parameters can be specified in the real-time feedback algorithm. As a first approximation,  $\beta \propto c_1 \cos(\omega_0\tau)$ and the linear growth rate $\nu_\text{wod}$ is positive when $\beta>\alpha$, which can be achieved  for a range of delays $\tau$ and amplification factor $c_1$.\\
Fig. \ref{fig:meas_tech}a shows the measurements of the eigenfrequency and growth/decay rate of the first transversal mode when the electroacoustic feedback is active, for different feedback delays $\tau$ and for fixed $c_1$ and $c_2$. When the electroacoustic system is linearly unstable (red background), growth rate measurements were performed by fitting exponential curves on the transient growth of the acoustic amplitude (Fig. \ref{fig:meas_tech}b) and do the average over 10 realizations. The standard deviation to mean ratio of such growth rate measurements can be seen Fig. \ref{fig:stablim_tech}b for different values of $c_1$, which shows very good accuracy. For the linearly stable cases (green background), the measurement technique is the same as for measuring the natural damping of the mode (Fig. \ref{fig:meas_tech}c and \ref{fig:meas_tech}d), namely performing sweep measurements. The accuracy of the sweep measurements was checked using decay rate  measurements: the second loudspeaker is used to excite the system at its resonance frequency, and when the excitation is stopped the transient decay of the acoustic amplitude is fitted with an exponential curve. These decay rate measurements are averaged over 10 realizations and gives similar accuracy as the growth rate measurements. The decay rates $\nu_\text{wod}$ obtained from the sweep measurement was always within one standard deviation compared to the one obtained from decay rate measurement.\\
In the remainder of this paper, the delay $\tau=2.2$ ms, which gives the most constructive feedback, and the saturation coefficient $c_2$ are kept constant. The linear growth rate $\nu_\text{wod}$ of the system without dampers is varied by adjusting $c_1$, according to the linear regression shown in Fig. \ref{fig:stablim_tech}a.
The effective saturation constant $\kappa$ can be deduced from the square root evolution of the oscillation amplitude when $\nu_\text{wod}$ is varied. Indeed, the theoretical limit cycle amplitude of eq. \eqref{eq:VdP} is $A_\eta=\sqrt{-8\nu_\text{wod}/\kappa}$ (see \cite{noiray2013ijnlm}), and  $\kappa = -0.08$ s\textsuperscript{-1}Pa\textsuperscript{-2} was deduced from these measurements.

\subsection{Addition of dampers}

\begin{figure}[h]
\centering
\includegraphics[width=\figswidth]{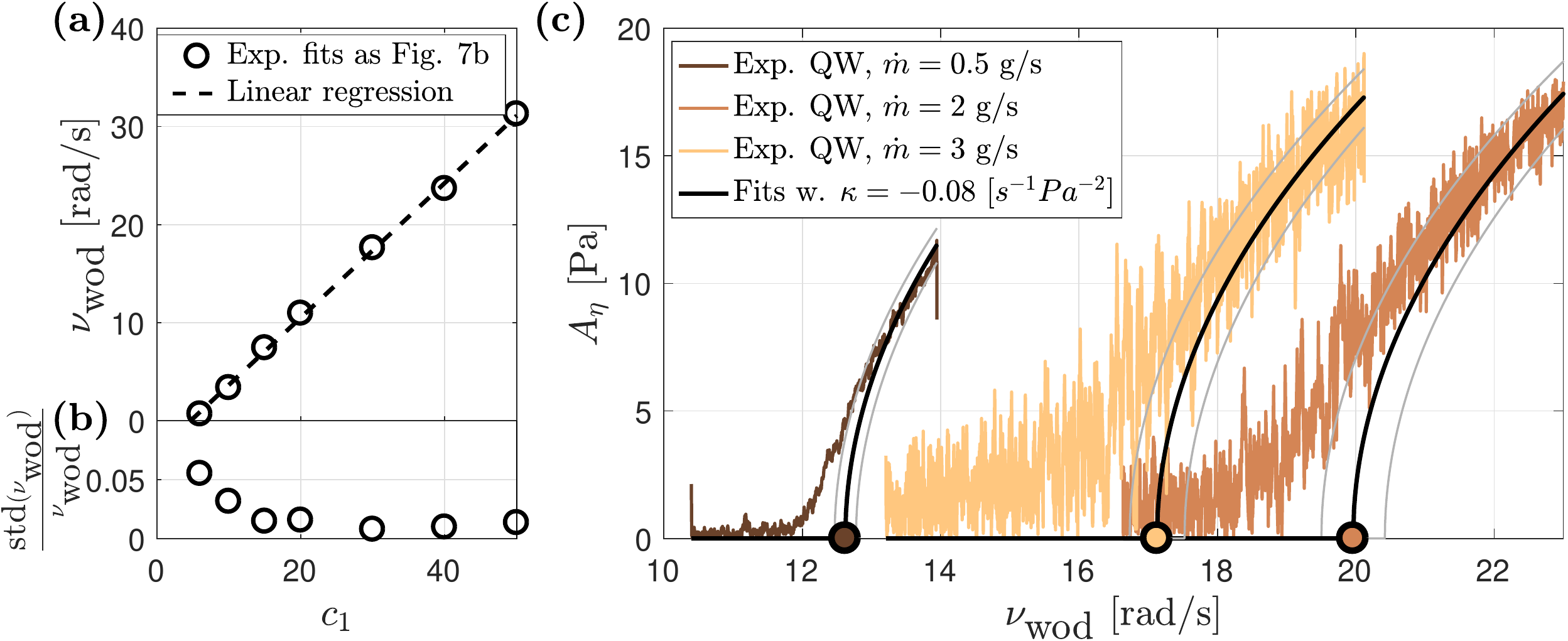}
\caption{\textbf{(a)} Relation between $c_1$ and $\nu_\text{wod}$ for several growth rate measurements, and linear regression verifying proportionality. \textbf{(b)} Standard deviation to mean ratio for the previous growth rate measurements, showing very good accuracy. \textbf{(c)} Example of the stability limit measurements for the tuned QW for 3 different mass flows. $\nu_\text{wod}$ is slowly ramped down and the resulting pressure envelope is fitted using $\kappa=0.08$ [s\textsuperscript{-1}.Pa\textsuperscript{-2}].}
\label{fig:stablim_tech}
\end{figure}

\begin{figure}[h]
\centering
\includegraphics[width=\figswidth]{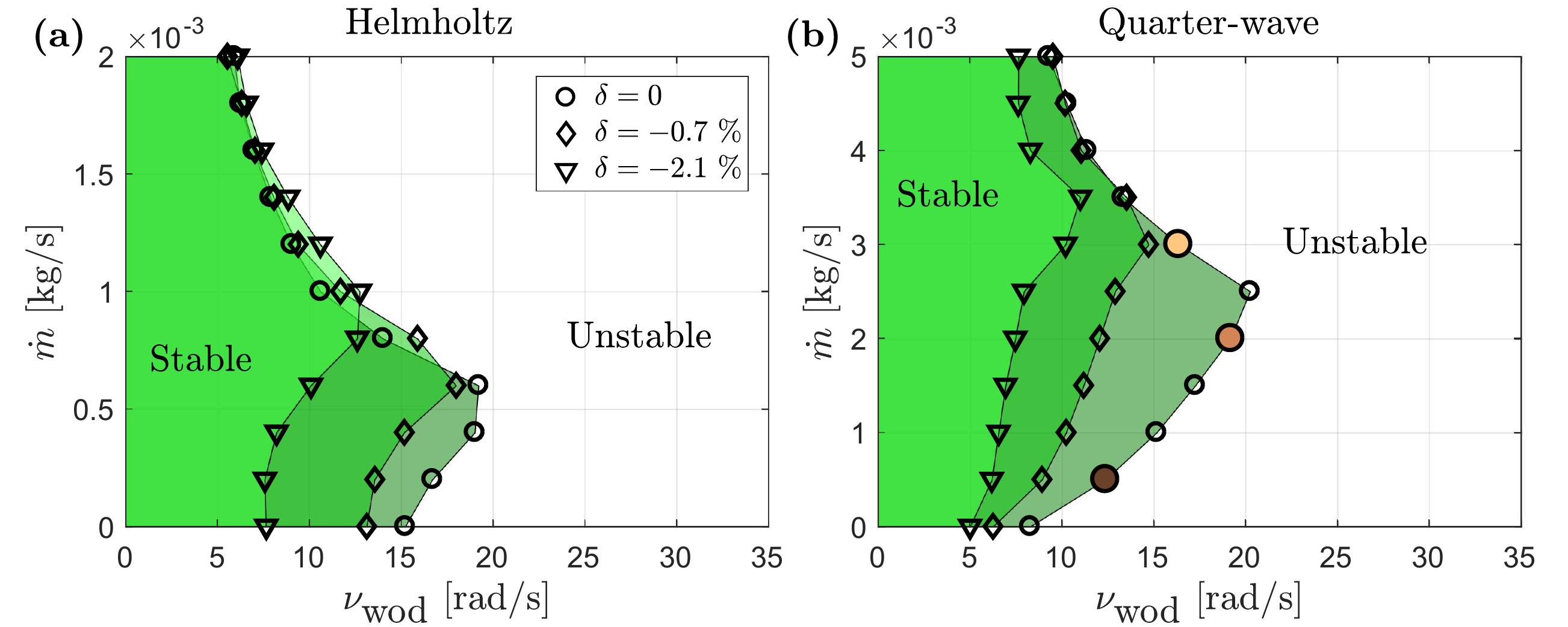}
\caption{Experimental stability limits as function of purge mass flow $\dot{m}$ and mode growth rate $\nu_\text{wod}$ for the perfectly tuned case and two detuned cases with $\delta = (\omega_{H,Q} - \omega_0)/\omega_0 = -0.7$ and $-2.1$ \%. \textbf{(a)} HH, \textbf{(b)} QW. The three colored points correspond to the fits in Fig. \ref{fig:stablim_tech}c.}
\label{fig:stablim_exp}
\end{figure}

\noindent A HH or QW damper can be connected to the cavity and fed with a purge flow as shown in Fig. \ref{fig:cavity_setup}. For the stability measurements, the mass flow $\dot{m}$ is varied between 0 and 2 g/s for the HH, and between 0 and 5 g/s for the QW. Without damper, the stability of the first transversal mode depends on the sign of $\nu_\text{wod}$: if it is positive, the mode is linearly unstable, if it is negative, the mode is linearly stable. With damper, the damping of the system is increased, and the stability limits change. These stability limits  depend on the feedback loop gain $c_1$  and on the mass flow of the purge air going through the damper. The determination of these new limits is done as follows: for each tuned damper (HH and QW) and each mass flow, the gain $c_1$ is slowly ramped down, starting from a value where the system with dampers is linearly unstable and is on a limit cycle. The linear growth rate $\nu_\text{wod}$ is a linear function of $c_1$ as shown in Fig. \ref{fig:stablim_tech}a, and the  ramping down of $c_1$ is equivalent to a 1 rad/s decrease of $\nu_\text{wod}$ in 30 second. The  decrease of the bifurcation parameter $\nu_\text{wod}$ is therefore quasi-steady and one can fit the corresponding acoustic envelope using a function that is proportional to $\sqrt{\nu_\text{wod}}$. The origin of that fit is the bifurcation point, which defines the stability limit. The results of three of those measurements and the respective fits can be seen in Fig. \ref{fig:stablim_tech}c. The error was estimated by doing a fit on the truncated acoustic envelope curve, starting from the part with the highest amplitude. As shown in Fig. \ref{fig:stablim_tech}c, the confidence intervals are very small compared to the variation caused by mass flow variation.\\
The length of the back-volume of the HH and the length of the QW are then shortened to obtain a detuning $\delta = (\omega_{H,Q} - \omega_0)/\omega_0 = -0.7$ and $-2.1$ \%, and the measurements are repeated. The stability limits obtained by employing this procedure, for tuned and detuned dampers, are presented in Fig. \ref{fig:stablim_exp}. As expected the zone of stability shrinks when the damper is detuned. Those results will be compared with the model later on.

\begin{figure}[h]
\centering
\includegraphics[width=0.6\figswidth]{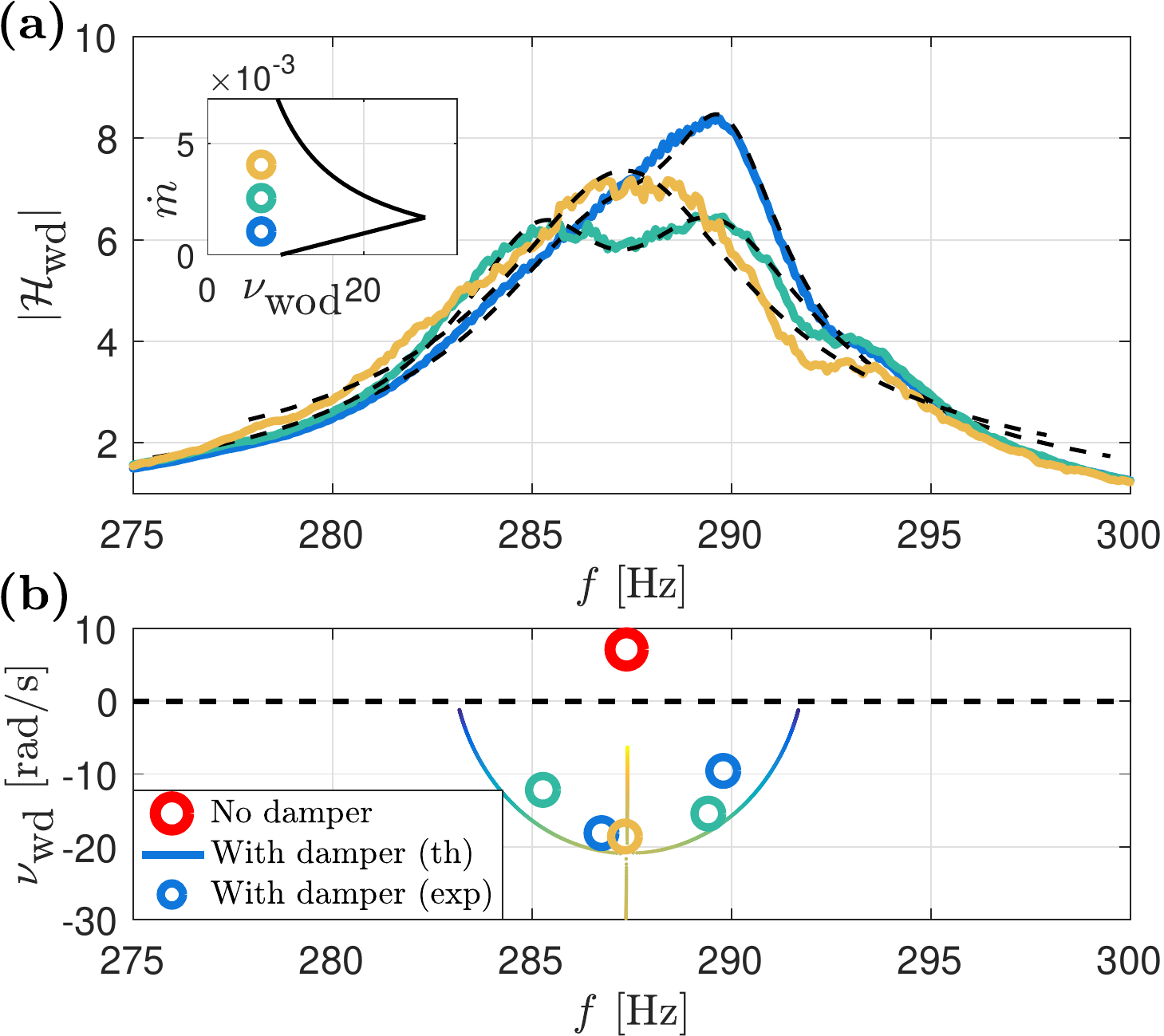}
\caption{Evolution of the experimental spectrum \textbf{(a)} and of the corresponding roots \textbf{(b)} of the cavity with $\nu_\text{wod}=7$ rad/s by addition of a tuned QW. Increasing the mass flow also influences the detuning (through the length correction) and the damper can only be tuned for a certain mass flow ($\dot{m}=2.5$ g/s in this case).}
\label{fig:comp_model_exp_QWtuned}
\end{figure}

\noindent The feedback gain $c_1$ is now fixed such that $\nu_\text{wod}=7$ rad/s. The eigenvalues of the first transversal mode of the chamber-damper coupled system are determined as function of the purge mass flow going through the damper. Since all of those measurements take place inside the stable zone, the eigenvalues are obtained using sweep measurements as done in Fig. \ref{fig:meas_tech}c and d.\\
When a damper with low damping is added to the cavity, mode splitting occurs. This was demonstrated experimentally in e.g. \cite{pietrzko2008ast,yu2008jasa,soon2012ksnve,li2007jasa}. In the detuned case, one mode is much closer to the stability limit than the other one, and dominates the frequency response of the transfer function. In that case, a 2\textsuperscript{nd} order transfer function fit yields satisfactory results. In the tuned case however, both modes should have equal decay rates but different frequencies for low purge mass flow. This means that the experimental transfer function should be fitted by a 4\textsuperscript{th} order transfer function to capture both modes correctly. In practice, however, since the mass flow through the damper also influences the end correction and thus the detuning of the damper, symmetric mode splitting can only be achieved at one particular mass flow as one can see in Fig. \ref{fig:comp_model_exp_QWtuned} for the tuned QW. This is even worse for the HH, since it is much more prone to mass flow-induced detuning (see Fig. \ref{fig:R_coeff}a). Therefore, only a 2\textsuperscript{nd} order transfer function is used to determine the poles of the transfer function in a systematic manner, accepting that this technique might lead to small errors around the mass flow for which symmetric mode splitting occurs. 

\begin{figure}[h]
\centering
\includegraphics[width=\figswidth]{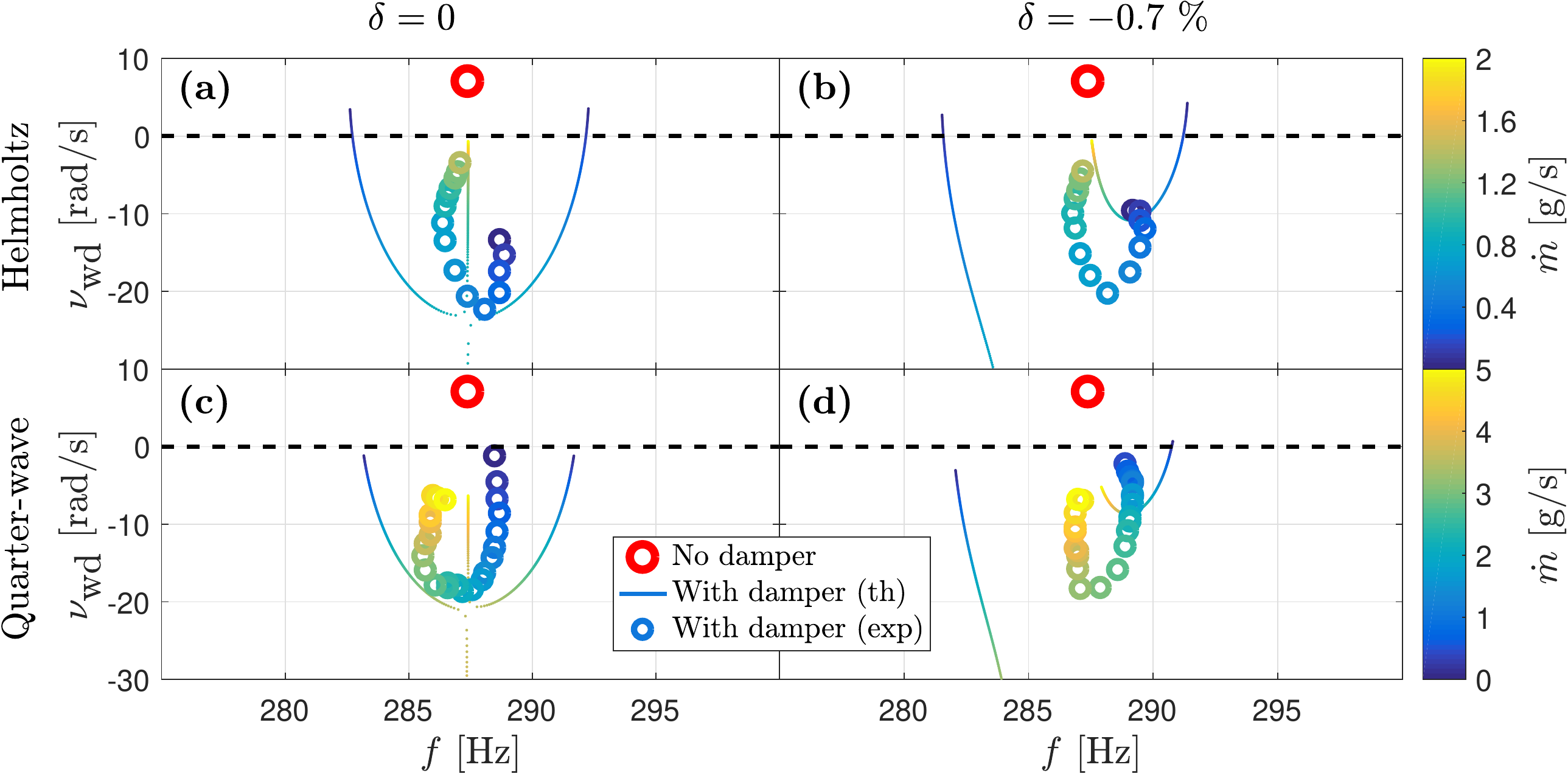}
\caption{Evolution of an unstable eigenmode (red circle) of the cavity with $\nu_\text{wod}=7$ rad/s by addition of dampers according to the model (continuous curve) and to the experiments (circles) for different mass flows (color scale) for the tuned \textbf{(a)} and detuned \textbf{(b)} Helmholtz resonator and for the tuned \textbf{(c)} and detuned \textbf{(d)} Quarter-Wave resonator.}
\label{fig:comp_model_exp_nu7}
\end{figure}

\noindent The results of this fit compared with the predictions from the analytical model (which will be described in the next section) are shown in Fig \ref{fig:comp_model_exp_nu7} for both dampers, either tuned or slightly detuned ($\delta=-0.7$ \%). As expected, the model is not accurate for the HH damper at low mass flows: a nonlinear dissipation term would be needed to capture its behavior at low mass flow. The agreement between theory and experiments is otherwise good. Comparing HH and QW resonator, the HH damper achieves better stabilization for low mass flow than the QW for similar damper volume. The mass flow needed to achieve best stabilization is higher for the QW than for the HH. Even if both are used at their best mass flow condition, the HH damper achieves slightly better stabilization than the QW (which is also predicted by the model).

\FloatBarrier

\begin{figure}
\centering
\includegraphics[width=\figswidth]{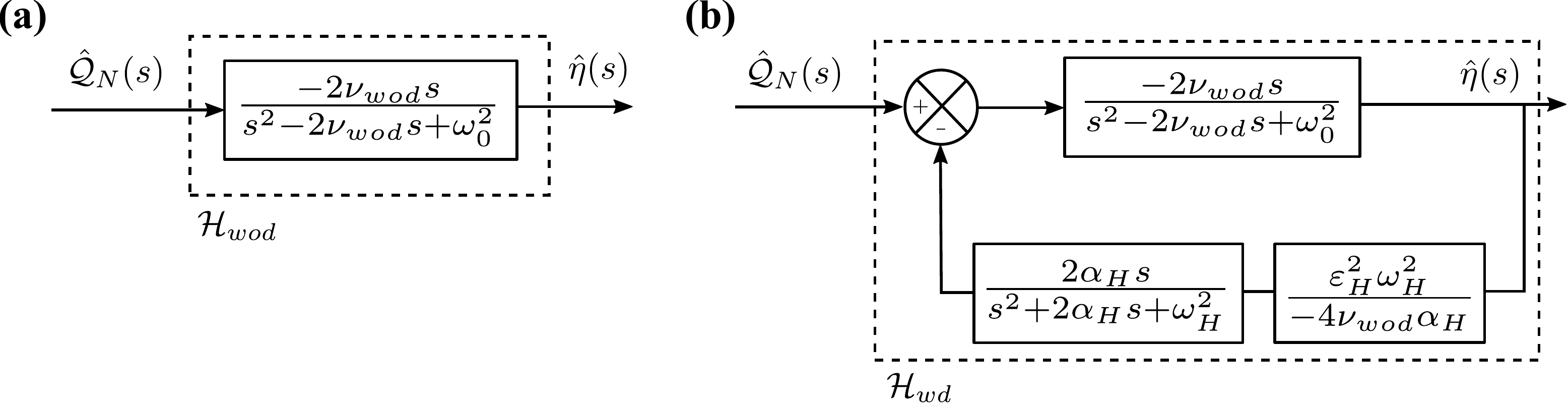}
\caption{Block diagram representation of the system \textbf{(a)} without and \textbf{(b)} with HH dampers as in \cite{noiray2012jsv}.}
\label{fig:blockdiagram}
\end{figure}

\section{Analytical model and optimal damping}

\subsection{Analytical model}
\label{subsec:anal_model}
\noindent The derivation of the model used for comparison with the experiments in Figs. \ref{fig:comp_model_exp_QWtuned} and \ref{fig:comp_model_exp_nu7} can be found in \ref{sec:appA} based on the work of \citep{noiray2012jsv}. With a single dominant mode, the pressure in the chamber can be approximated by $p(t, \mathbf{x})=\eta(t) \psi(\mathbf{x})$ with $\psi(\mathbf{x})$ the acoustic eigenmode and $\eta(t)$ its amplitude. The transfer function of the chamber-damper coupled system (Fig. \ref{fig:blockdiagram}b) can be written in the HH case as follows:
\begin{equation}
\mathcal{H}_{\text{wd}} (s) =\frac{\hat{\eta}(s)}{\hat{\mathcal{Q}}_N (s)} = \frac{-2 \nu_\text{wod} s \left( s^2 + 2\alpha_H s + \omega_{H}^2 \right)}{\left( s^2 -2 \nu_\text{wod} s + \omega_0^2 \right) \left( s^2 + 2\alpha_H s + \omega_{H}^2 \right) + s^2 \omega_{H}^2 \varepsilon_H^2},
\label{eq:transfer_func_H}
\end{equation}

\noindent with $\hat{\mathcal{Q}}_N (s)$ the noise component of the acoustic source in the chamber volume, $\omega_0$ the natural angular frequency of the dominant mode, $\nu_\text{wod}$ its growth/decay rate, $\omega_H$ the angular frequency of the damper and its damping $\alpha_H = R_H / 2\rho_d l$ and $R_H$ the resistive term from Eq. \ref{eq:res_H}. $\varepsilon_H$ is the damping efficiency factor defined as:

\begin{equation}
\varepsilon_H^2 = \frac{V_H}{V} \frac{\Psi_d}{\Lambda},
\end{equation} 

\noindent with $V_H=AL$ the volume of one damper, $V$ the chamber volume, $\Lambda$ the norm of the mode and $\Psi_d=\sum_{k=1}^{n} \psi^2(\mathbf{x}_k)$ a non-dimensional number describing the number and location of the dampers with respect to the pressure antinode of the mode. If $\varepsilon_H=0$, then $\mathcal{H}_{\text{wd}}=\mathcal{H}_{\text{wod}}$ shown in Fig. \ref{fig:blockdiagram}a. The equivalent expressions for the QW case can be found in \ref{sec:appA}. The previous description is  equivalent to the following time domain formulation:
\begin{equation}
	\left\{ \begin{array}{l}	
	\ddot{\eta} -2\nu_\text{wod}\dot{\eta} + \omega_0^2\eta = - \dfrac{\varepsilon_H^2 \omega_H^2 \rho l}{\Psi_d}\dot{u} \\ 
	\\
	 		\ddot{u} + 2\alpha_H\dot{u} + \omega_H^2 u =\dfrac{\Psi_d}{\rho l} \dot{\eta},
\end{array} \right.
\label{eq:coupsys_time}
\end{equation}
\noindent with $u$ the acoustic velocity in the damper neck. This system of two coupled ODEs can be expressed as:
\begin{equation}
	\dot{X}=MX \quad \text{with} \quad M=\left[\begin{array}{cccc} 2 \nu_\text{wod} & -\omega^2_0 & -\frac{\varepsilon^2_H \omega^2_H \rho l}{\Psi_d} & 0 \\ 1 & 0 & 0 & 0 \\ \frac{\Psi_d}{\rho l} & 0 & -2\alpha_H & -\omega^2_H \\ 0 & 0 & 1 & 0 \end{array}\right] \quad \text{and} \quad X=\left[\begin{array}{c} \eta \\ \dot{\eta} \\ u \\ \dot{u} \end{array}\right].
\label{eq:coupsys_matrix}
\end{equation}

\subsection{Optimal damping and exceptional points}
\label{subsec:anmod_results}
\noindent The analytical model presented in \ref{subsec:anal_model} is now used to determine the linear stability of the coupled system ``chamber-damper'', which was experimentally investigated in section \ref{sec:experiments}. To that end, one uses the geometrical and flow parameters which characterize the acoustic mode, the dampers and the coupling efficiency in this experimental configuration:

\begin{itemize}
\item The  first transversal mode of the cavity is considered, with $\omega_0\simeq 1803$ rad/s, $f_0=\omega_0/2 \pi\simeq$ 287 Hz. The electro-acoustic feedback allows to set $\nu_\text{wod}$ up to 35 rad/s.
\item The experimentally-identified linear relationships $R_H (\dot{m})$ and $R_Q (\dot{m})$ presented in   Fig. \ref{fig:zetafit} are used for the evaluation of the damping $2\alpha_H (\dot{m}) = R_H /\rho l$ and $2\alpha_Q(\dot{m})= R_Q /\rho L$. The effective lengths $l=l_p+l_\text{cor}$ and $L=L_p+L_\text{cor}$, for dampers connected to the large chamber, are not the same as in section \ref{subsec:R_meas} where they were connected to a duct. Therefore, the end corrections are determined again using the Helmholtz solver AVSP, for the ``chamber-damper'' arrangement,  which gives $l_{\text{cor}}=15.9$ mm and $L_{\text{cor}}=10.3$ mm.  The physical length $L$ is used for the tuning of the natural resonance frequency of the dampers $\omega_H$ and $\omega_Q$. 
\item   The damping efficiency factors $\varepsilon_H$ and $\varepsilon_Q$ depend on the damper-to-chamber volume ratio, on the   modal amplitude at the damper location and on the mode normalization factor. In the present configuration, $\Lambda=1/2$ and $\psi=0.95$ (see Fig. \ref{fig:cavity_setup}). 
\end{itemize}

\noindent The stability of the coupled system is obtained from Eq. \eqref{eq:transfer_func_H} for the range of purge mass flow $\dot{m}$ and linear growth rate $\nu_\text{wod}$ which were used in the experiments. In the perfectly tuned case with $\omega_0=\omega_H$, the stability limits of the coupled system for the HH damper case are explicitly obtained from the Routh-Hurwitz criterion:

\begin{equation}
\left\{ \begin{array}{l r}	
	\alpha_H \geq \nu_\text{wod}  & \text{RH1}\\  
	\\
	\omega_0^2 \varepsilon_H^2  \geq 4 \nu_\text{wod} \alpha_H  & \text{RH2}
\end{array} \right.
\label{eq:routh_H}
\end{equation}

\begin{figure}[h]
\centering
\includegraphics[width=\figswidth]{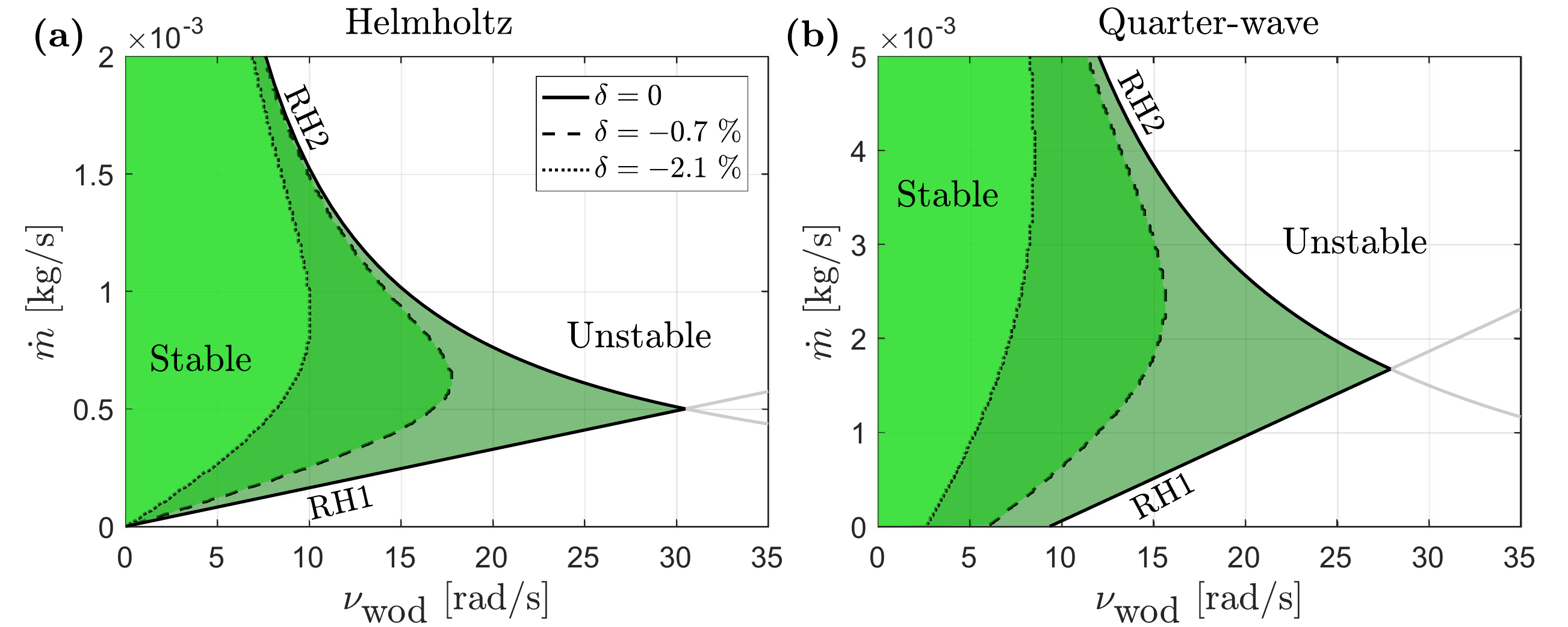}
\caption{Theoretical stability limits as function of purge mass flow $\dot{m}$ and growth rate $\nu_\text{wod}$ for the perfectly tuned case (Routh-Hurwitz criterion, analytical) and two detuned cases (numerical) with $\delta = (\omega_d - \omega_0)/\omega_0 = -0.7$ and $-2.1$ \%. \textbf{(a)} HH, \textbf{(b)} QW.}
\label{fig:stablim}
\end{figure}

\begin{figure}[h]
\centering
\includegraphics[width=0.5\figswidth]{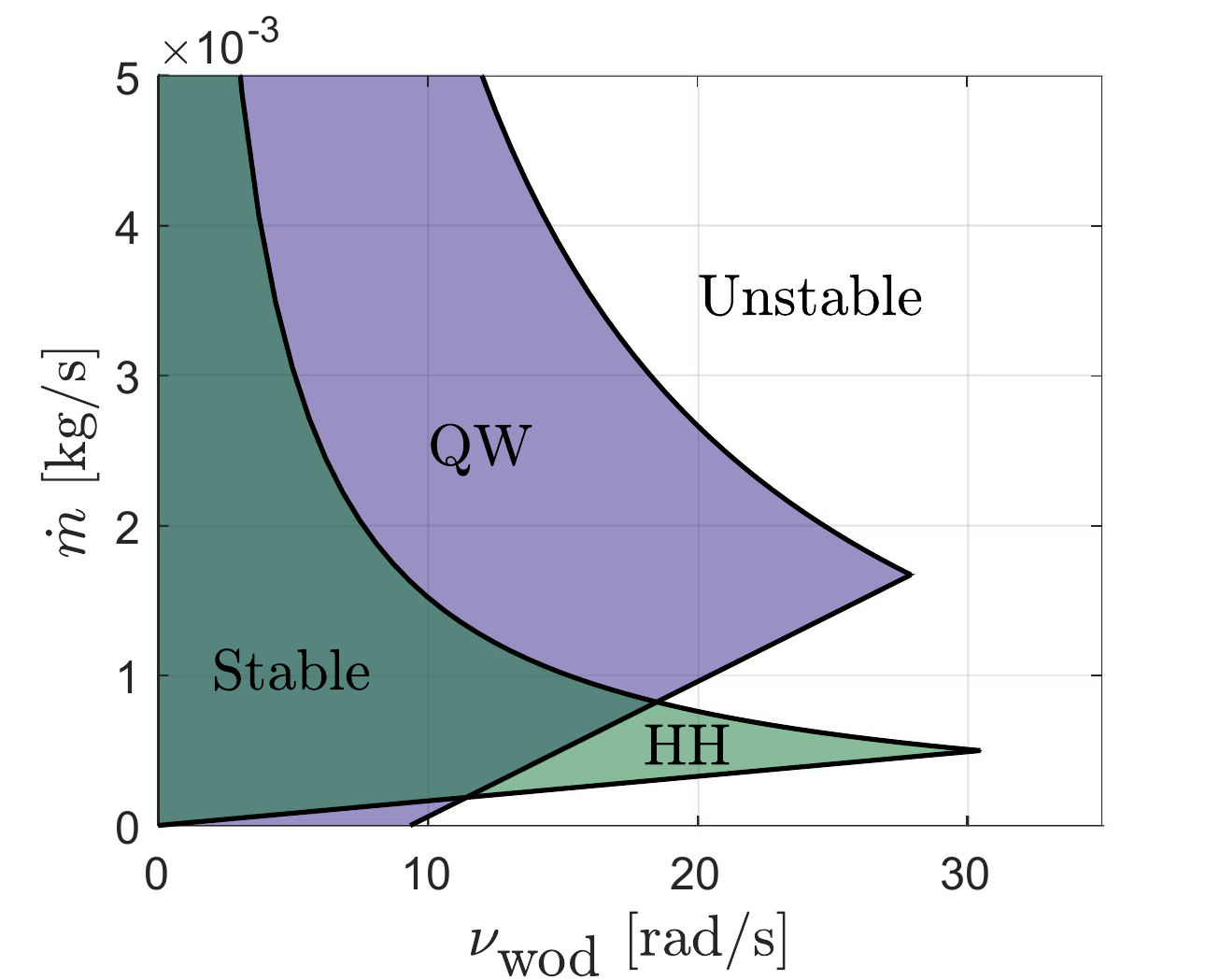}
\caption{Comparison of the theoretical stability limits as function of purge mass flow $\dot{m}$ and growth rate $\nu_\text{wod}$ for the perfectly tuned case between HH and QW.}
\label{fig:stablim_comp}
\end{figure}

\noindent The Routh-Hurwitz criterion can be easily translated as follows: the damping of the damper needs to be higher than the growth rate of the unstable mode, and the ratio weighting the feedback in the block diagram (Fig. \ref{fig:blockdiagram}b) needs to be greater than 1. For the QW damper case, the Routh-Hurwitz criterion is the same, replacing $\alpha_H$ by $\alpha_Q$ and $\varepsilon_H$ by $\varepsilon_Q$. For the detuned case, there is no analytical expression giving the linear stability boundaries. The stability limit is numerically determined by computing the poles of  Eq. \eqref{eq:transfer_func_H}, and searching the change of sign of the real part of the least stable of these poles. The theoretical stability limits for tuned and detuned Helmholtz and Quarter-Wave dampers are presented in Fig. \ref{fig:stablim} and are in good agreement with the ones measured experimentally (see Fig. \ref{fig:stablim_exp}). The comparison between the theoretical stability limits of the tuned HH and tuned QW (having quasi-identical volume) are shown in Fig. \ref{fig:stablim_comp}.

\begin{figure}[h]
\centering
\includegraphics[width=\figswidth]{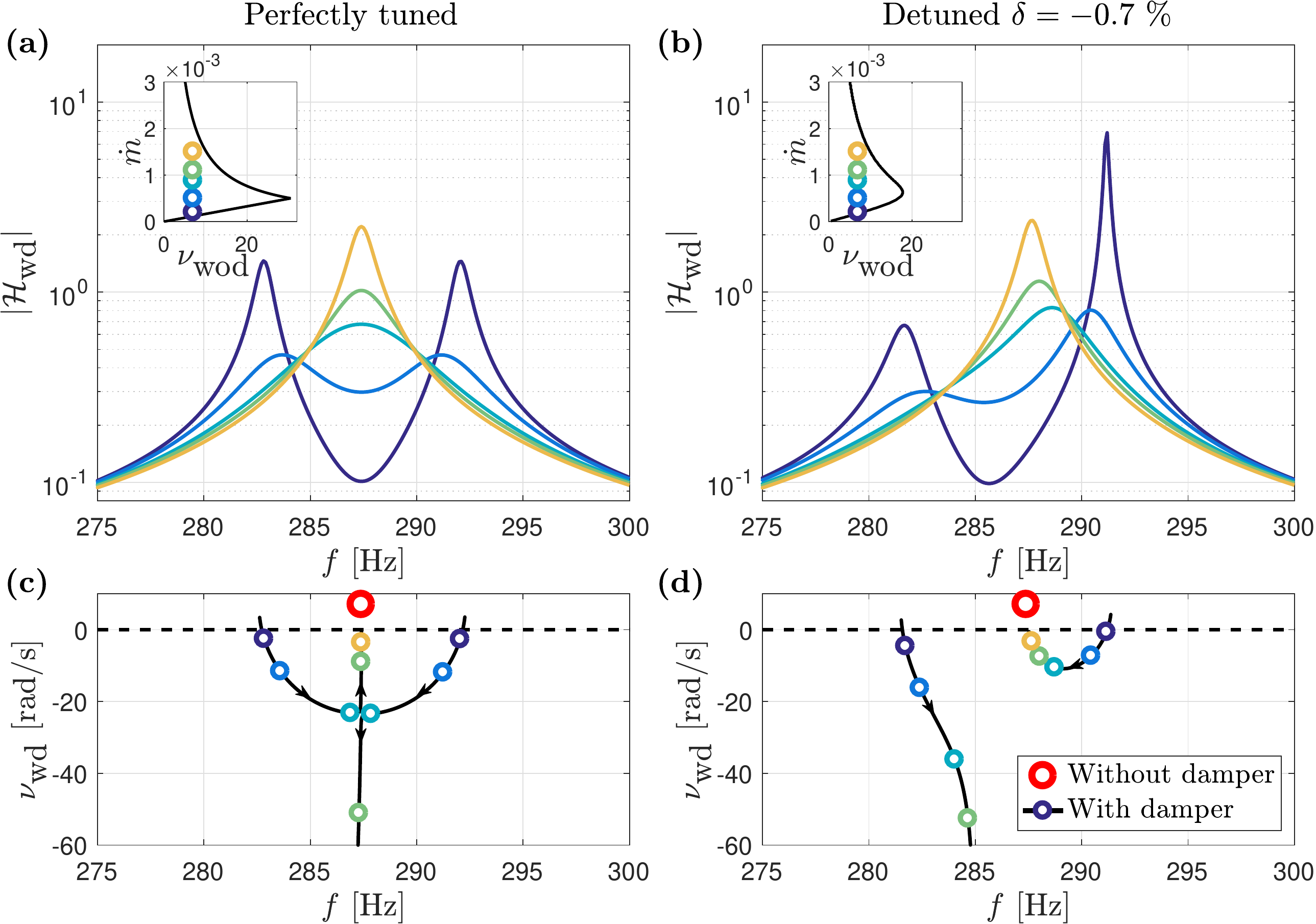}
\caption{Influence of increasing mass flow on the magnitude (top) and on the poles (bottom) of the transfer function describing the coupled system ``chamber-damper'' for an unstable mode with $\nu_\text{wod} = 7$ rad/s and HH resonator. \textbf{(a)} and \textbf{(c)} tuned, \textbf{(b)} and \textbf{(d)} detuned. For the tuned case, the pair of eigenvalues coalesce at the exceptional point for a specific mass flow $\dot{m}$, which corresponds to the most linearly stable coupled system.}
\label{fig:spec_ev}
\end{figure}

\begin{figure}[h]
\centering
\includegraphics[width=\figswidth]{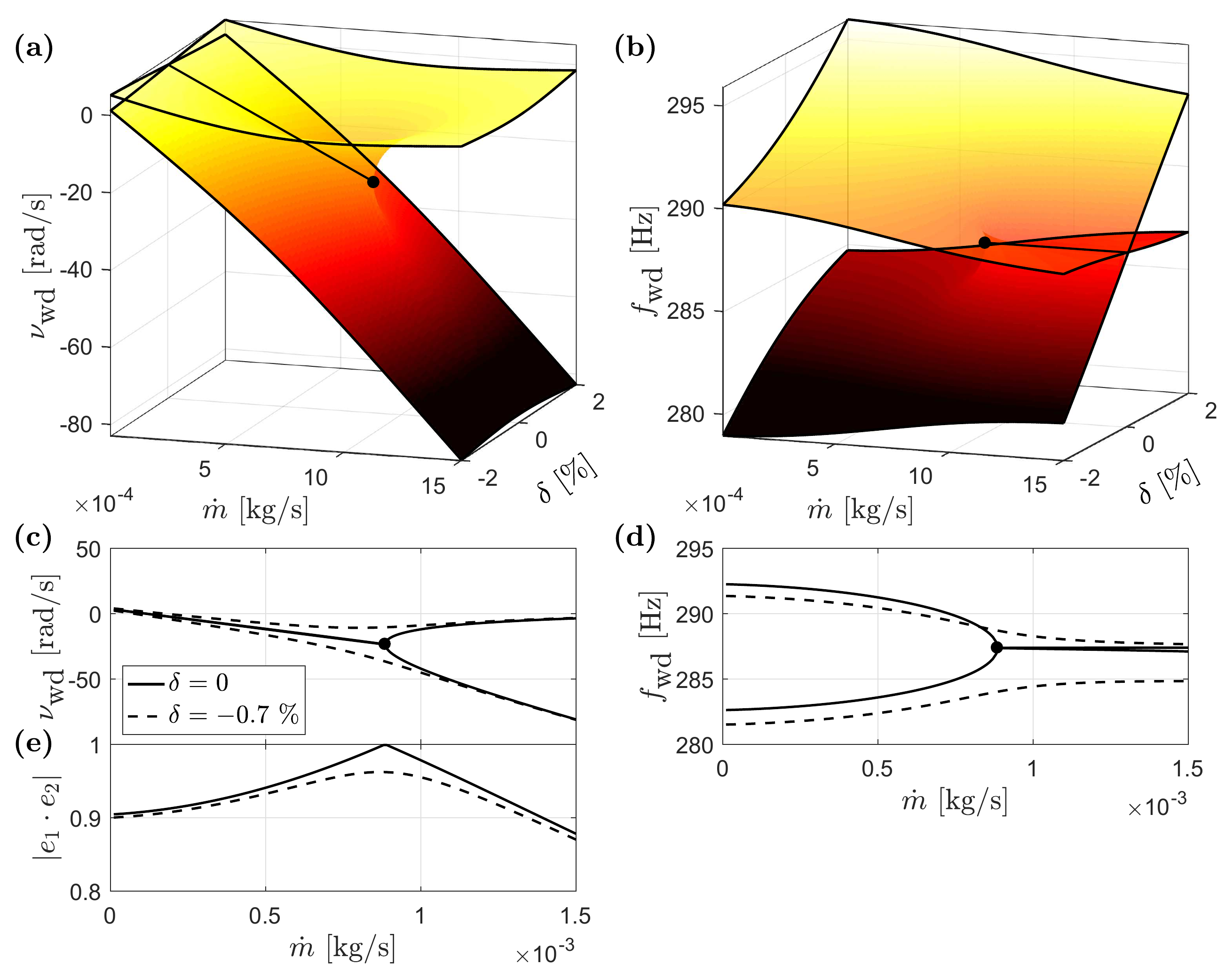}
\caption{Loci of the coupled system eigenvalues around the EP for $\nu_\text{wod} = 7$ rad/s. \textbf{(a)} $\Re(\lambda)=\nu_\text{wd}$ [rad/s] as function of purge mass flow $\dot{m}$ and detuning $\delta$  and \textbf{(c)} cuts of the surface for $\delta=$ 0 and -2 Hz, \textbf{(b)} $\Im(\lambda)/2\pi=f_\text{wd}$ [Hz] as function of purge mass flow $\dot{m}$ and detuning $\delta$, \textbf{(d)} cuts of the surface for $\delta=$ 0 and -2 Hz. The EP is represented by the black dot. \textbf{(e)} Scalar product of the coupled system normalized eigenvectors $e_1$ and $e_2$, showing their coalescence at the EP.}
\label{fig:EP_nappes}
\end{figure}

\noindent Fig. \ref{fig:spec_ev} shows the influence of increasing purge mass flow $\dot{m}$ on the magnitude and on the poles of the transfer function characterizing the coupled system ``chamber-damper'' for fixed $\nu_\text{wod}=7$ rad/s. In both cases, the inset shows the coordinates $(\nu_\text{wod};\dot{m})$ in the stability diagram, which are considered for this analysis. For situations where the damper is tuned to the eigenfrequency of the system without dampers ($\omega_0=\omega_H$ or $\omega_0=\omega_Q$), the pair of eigenvalues of the system with damper merge at a so-called \emph{exceptional point} (EP) \cite{seyranian2005jpa}, when a critical purge mass flow $\dot{m}_\text{EP}$ through the damper is reached. For $\dot{m}<\dot{m}_\text{EP}$, the poles of the system  with the damper symmetrically split, with identical linear growth rate $\nu_\text{wd}$ and different frequencies $f_\text{wd}=f_0\pm\Delta f$. The associated transient dynamics is a decay of the oscillation amplitude with time constant $1/\nu_\text{wd}$, which is accompanied with a low-frequency amplitude beating of period $1/\Delta f$ \cite{ryu2015pre}. For $\dot{m}>\dot{m}_\text{EP}$, the pair of eigenvalues originating from the exceptional point exhibit the same frequency as the natural eigenfrequency of the system without damper ($\omega_\text{wd}=\omega_0$), but one of these eigenvalues has a larger linear growth rate $\nu_\text{wd}$ than the other and than the one at the EP, i.e. the associated eigenmode is less stable.\\
By increasing the purge mass flow, the best stabilization of the mode is therefore achieved when the eigenvalues and associated eigenmodes of the coupled system coalesce at the EP. This can be verified by computing the eigenvalues and eigenvectors of $M$ in Eq. \ref{eq:coupsys_matrix}: the scalar product between the normalized eigenvectors $e_1$ and $e_2$ corresponding to positive frequency can be seen in Fig. \ref{fig:EP_nappes}e. At the EP, the scalar product is 1, meaning that they coalesce. If one further increases the purge mass flow, the eigenvalues split into two separate modes at same frequency but different decay rates, one of them being stabilized and the other one destabilized by a further increase of the mass flow. Note that the mass flow giving the best mode stabilization is higher than the one minimizing the infinite norm of the frequency response. In the situation where the dampers are not well tuned (Fig. \ref{fig:spec_ev}b and \ref{fig:spec_ev}d), there is an avoided crossing of the eigenvalues and the system does not exhibit any EP when the purge mass flow is varied. In the present configuration, a detuning of 0.7 \% leads to a substantial root loci deviation and to an avoided crossing compared to the tuned damper scenario. The theoretical behavior presented in Fig. \ref{fig:spec_ev} is again in good agreement with the experimental measurements shown in Fig. \ref{fig:comp_model_exp_QWtuned} and Fig. \ref{fig:comp_model_exp_nu7}.\\

\noindent To complement this analysis, the eigenvalues $\lambda=\nu_\text{wd}+i\omega_\text{wd}$ of the coupled system ``chamber-damper''  are presented in the form of Riemann sheets as function of the damper detuning  $\delta$ and of the damper purge mass flow $\dot{m}$ in Fig. \ref{fig:EP_nappes} for $\nu_\text{wod}=7$ rad/s. In Fig. \ref{fig:EP_nappes}c, one can see that the EP is the point where the real part of the double root of the quartic polynomial \cite{rees1922amm} (denominator of Eq. \eqref{eq:transfer_func_H}) is minimum. When $\omega_H=\omega_0$ (tuned damper), one can explicitly deduce the associated HH resonator damping:

\begin{equation}
\alpha_{H,\text{EP}} = \omega_0 \varepsilon_H - \nu_\text{wod}. 
\label{eq:bestgamma}
\end{equation}

\noindent Klaus and co-workers \cite{klaus2014aa} recently drawn a similar result in the context of room acoustics. In that study, dissipative resonators are employed to minimize reverberation time, which means that, in contrast with the present work, the acoustic enclosure is already linearly stable before the implementation of the damper, which are used to further stabilize it. Here, the damping coefficient is linearly related to the mass flow through the damper and Eq. \eqref{eq:bestgamma} corresponds to

\begin{equation}
\dot{m}_\text{EP} (\nu_\text{wod}) = \dfrac{\rho a l}{\zeta_H}(2 \omega_0 \varepsilon_H - 2 \nu_\text{wod}).
\label{eq:bestmdot}
\end{equation}

\begin{figure}[h]
\centering
\includegraphics[width=0.75\figswidth]{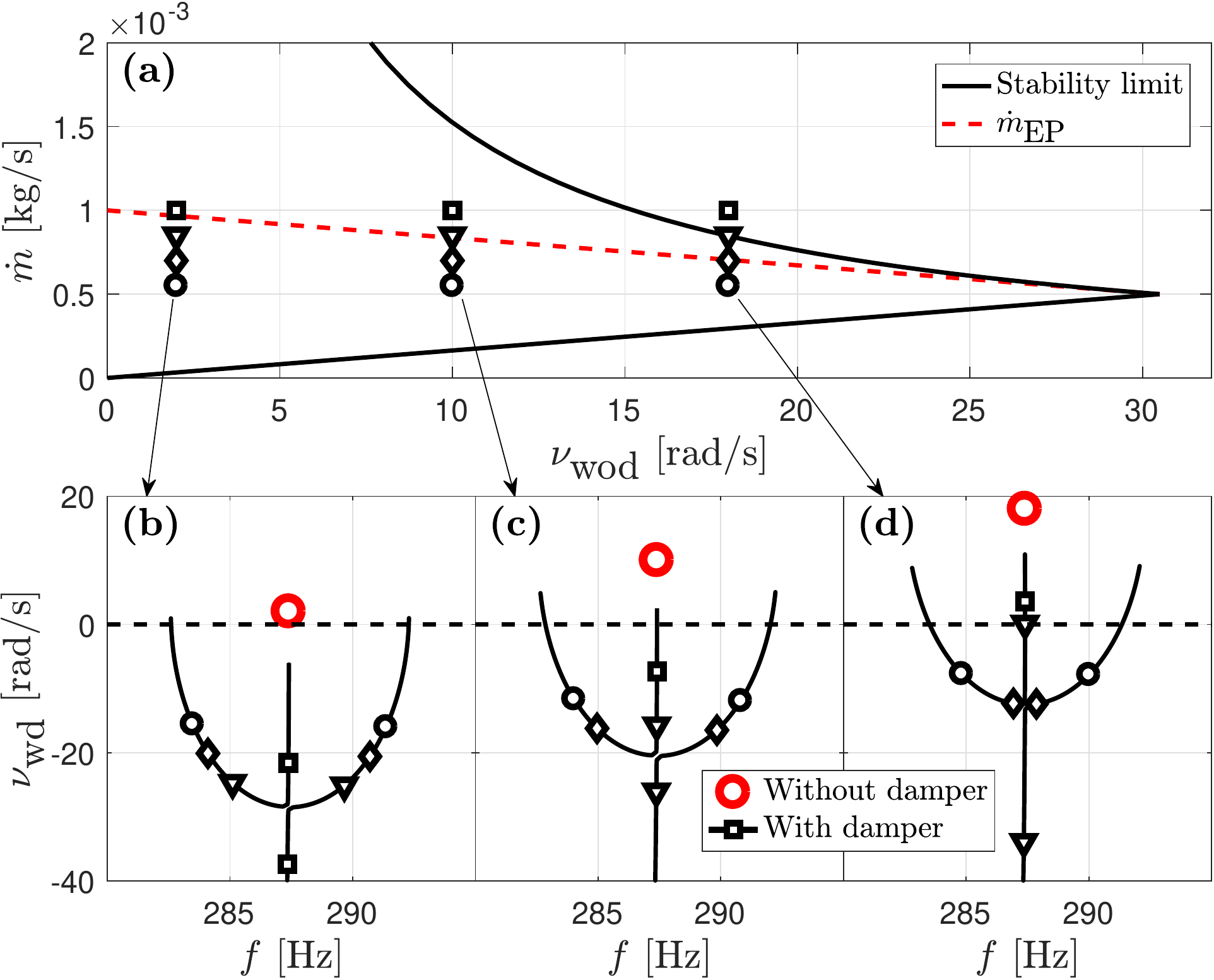}
\caption{Influence of growth rate $\nu_\text{wod}$ on the mass flow $\dot{m}_\text{EP}$ at which the exceptional point (and thus the best stabilization) is achieved. The dashed red line corresponds to Eq. \ref{eq:bestmdot}}
\label{fig:best_mdot_intro}
\end{figure}

\noindent A similar expression can be derived for the QW damper and includes the acoustic boundary layer losses. The location of the EP in the linear stability diagram is shown in Fig. \ref{fig:best_mdot_intro} as a dashed red line.

\begin{figure}[h]
\centering
\includegraphics[width=\figswidth]{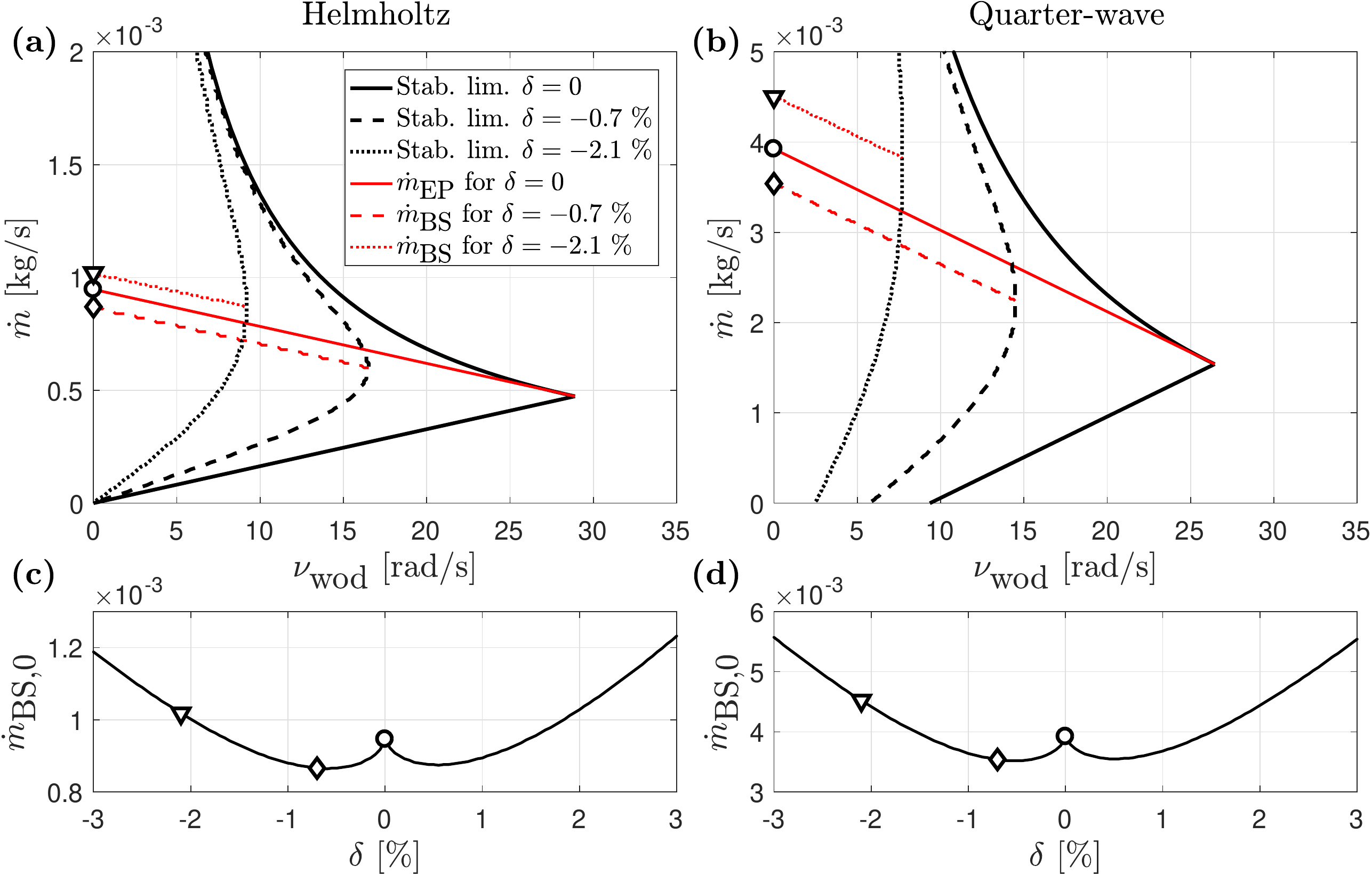}
\caption{Influence of detuning on the mass flow giving best stabilization. \textbf{(a)} $\dot{m}_\text{BS}$ as a function of $\nu_\text{wod}$ within stability limit for 3 different detuning values. The plain red curve corresponds to $\dot{m}_\text{EP}$ (Eq. \ref{eq:bestmdot}) whereas the dashed and dotted red curves were obtained numerically. The slope stays the same but the y-intercept varies. \textbf{(c)} y-intercept of $\dot{m}_\text{BS}$ as a function of the detuning $\delta$. \textbf{(b)} and \textbf{(d)}: same for the QW damper.}
\label{fig:stablim_bestmdot}
\end{figure}

\noindent The mass flow giving the best stabilization for detuned dampers is obtained numerically, and the results are shown in Fig. \ref{fig:stablim_bestmdot}. One can see that the slope of the lines indicating the maximum damping does not depend on the detuning.

\noindent For configurations with only one type of dampers, which can be a practical requirement in order to avoid having to manufacture, test and validate several geometries of dampers, the present work shows that the optimum damping is reached when the purge mass flow is adjusted close to the EP of the coupled system “chamber-dampers”. 
Combining dampers of different geometries, which address the same dominant acoustic mode, is also possible. In that case, the number of parameters in the system increases and the simple analytical expression given in eq. (24) for the optimum damper mass flow cannot be used. Still, a numerical optimization could be performed to find the mass flows for each of the dampers, which lead to the optimum modal damping.   
\\
\noindent Note that exceptional points exhibit an extreme sensitivity to parameter variation, which can be showed analytically. Let us note $D(s,\delta,\alpha_H)$ the polynomial corresponding to the denominator of Eq. \ref{eq:transfer_func_H}, whose roots $\lambda$ are the eigenvalues of the chamber-damper coupled system.

\begin{small}
\begin{equation}
D(s,\delta,\alpha_H) =\left( s^2 -2 \nu_\text{wod} s + \omega_0^2 \right) \left( s^2 + 2\alpha_H s + \omega_0^2 (1+\delta)^2 \right) + s^2 \omega_0^4 (1+\delta)^2 \varepsilon_H^2.
\label{eq:D_def}
\end{equation}
\end{small}

\noindent At the EP, the eigenvalues are 2 complex conjugate double roots $\lambda_\text{EP}=\nu_\text{wd,EP}\pm i\,\omega_\text{wd,EP}$. One can identify $D(s,0,\alpha_{H,\text{EP}})=(s-\lambda_\text{EP})^2(s-\lambda^*_\text{EP})^2$, with $\alpha_{H,\text{EP}}$ from Eq. \ref{eq:bestgamma}, giving

\begin{equation}
\nu_\text{wd,EP}=\nu_\text{wod} - \frac{\omega_0 \varepsilon_H}{2}, \qquad \omega_\text{wd,EP}=\sqrt{\omega_0^2-\nu_\text{wd,EP}^2}.
\label{eq:lambdaEP}
\end{equation}
\noindent Using Eqs. \eqref{eq:D_def} and \eqref{eq:lambdaEP}, one can show that the partial derivative of $D$ with respect to $s$ vanishes at the EP, i.e. $\partial_s D=0$ for $s=\lambda_\text{EP},\,\delta=0,\,\alpha_H=\alpha_{H,\text{EP}}$. One can also get analytical expressions of the other partial derivatives around the EP as function of $\omega_0$, $\nu_\text{wod}$ and $\varepsilon_H$. This gives the following Taylor expansion for a root $\lambda$ of the polynomial $D$ around the EP:
\begin{small}
\begin{equation}
0=D(\lambda,\delta,\alpha_H)= \frac{1}{2} (\lambda-\lambda_\text{EP})^2 \left.\dfrac{\partial^2 D}{\partial s^2}\right|_\text{EP} + \delta \left.\dfrac{\partial D}{\partial \delta}\right|_\text{EP} + (\alpha_H - \alpha_{H,\text{EP}}) \left.\dfrac{\partial D}{\partial \alpha_H}\right|_\text{EP}+...
\label{eq:EP_sensitivity}
\end{equation}
\end{small}

\noindent With $\delta=0$ the following approximation can be derived for a root $\lambda$ around the EP:
 
 \begin{small}
\begin{equation}
\lambda\simeq\lambda_\text{EP} +\left( 2\dfrac{\partial_{\alpha_H} D(\lambda_\text{EP},0,\alpha_{H,\text{EP}})}{\partial_{ss} D(\lambda_\text{EP},0,\alpha_{H,\text{EP}})} \right)^{1/2}\, \sqrt{\alpha_H-\alpha_{H,\text{EP}}}
\label{eq:Puiseux_series1}
\end{equation}
\end{small}

\noindent Similarly, if $\alpha_H=\alpha_{H,\text{EP}}$ the following approximation can be derived for a root $\lambda$ around the EP:

 \begin{small}
\begin{equation}
\lambda\simeq\lambda_\text{EP} +\left( 2\dfrac{\partial_{\delta} D(\lambda_\text{EP},0,\alpha_{H,\text{EP}})}{\partial_{ss} D(\lambda_\text{EP},0,\alpha_{H,\text{EP}})} \right)^{1/2}\, \sqrt{\delta}
\label{eq:Puiseux_series2}
\end{equation}
\end{small}
 
\noindent Eqs. \ref{eq:Puiseux_series1} and \ref{eq:Puiseux_series2} are Puiseux series \cite{seyranian2003book,kato2013book}. For Eq. \ref{eq:Puiseux_series2}, the sensitivity $d\lambda/d\delta \propto 1/\sqrt{\delta}$, which tends to infinity when $\delta \rightarrow 0$. A similar dependency can be found when $\alpha_H \rightarrow \alpha_{H,\text{EP}}$, thus showing the infinite sensitivity to parameter variation around the EP.
\FloatBarrier
\begin{figure}[h]
\centering
\includegraphics[width=\figswidth]{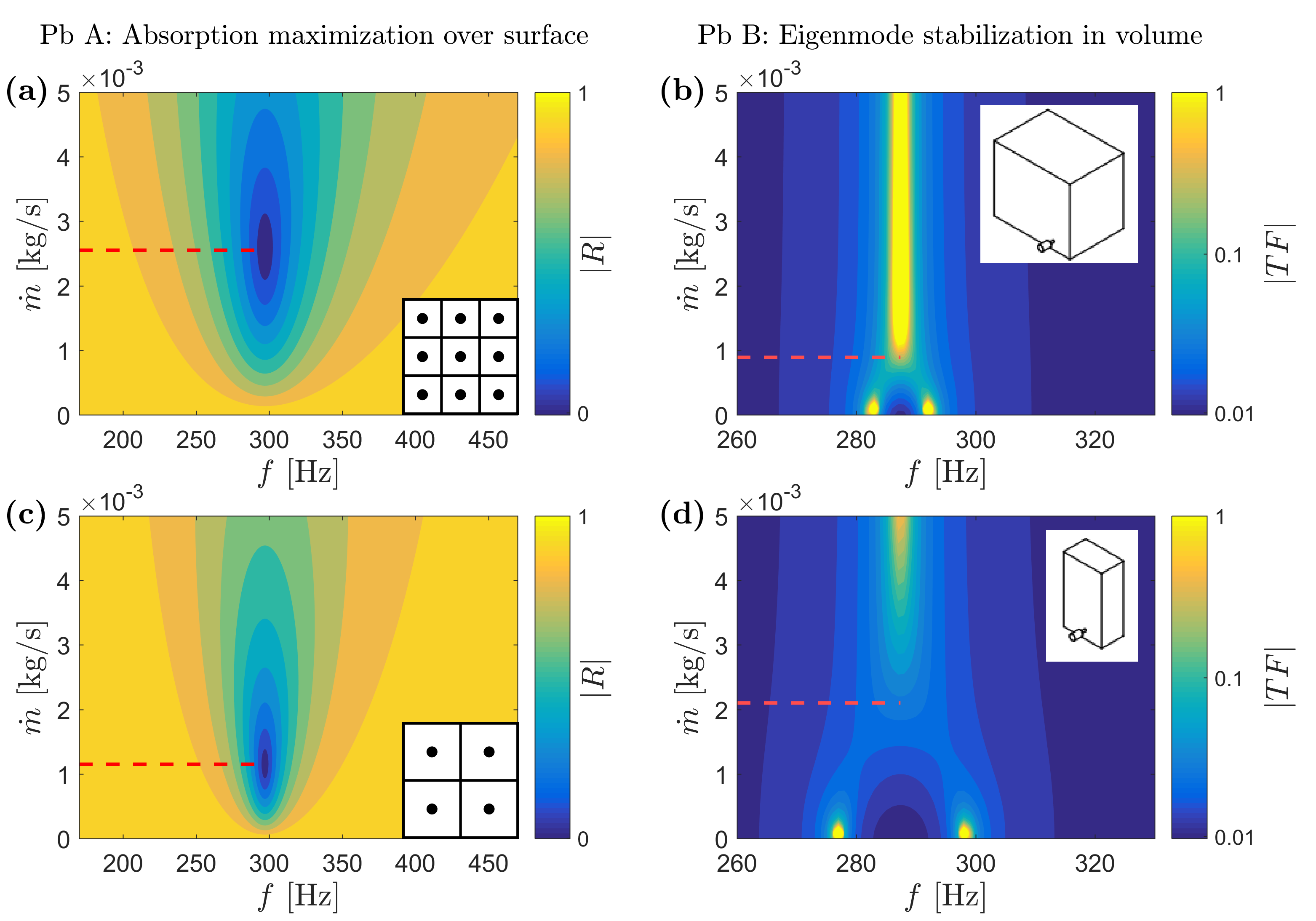}
\caption{For the exact same damper, comparison between reflection coefficient absolute value \textbf{(a)} for porosity $\sigma = 0.055$ and \textbf{(c)} for porosity $\sigma = 0.024$, and coupled damper-cavity spectrum \textbf{(b)} for efficiency factor $\varepsilon_H^2 = 0.0011$ ($V_c=0.2$ m\textsuperscript{3}) and \textbf{(d)} for efficiency factor $\varepsilon_H^2 = 0.0056$ ($V_c=0.041$ m\textsuperscript{3})}
\label{fig:bestmdot_RvsGRR}
\end{figure}

\section{Conclusion}

\noindent The optimization of the damper purge mass flow has been done in two different setups throughout this paper: either for minimizing the reflection coefficient at resonance in an impedance tube (section \ref{subsec:R_meas}) or for achieving the best stabilization of an unstable eigenmode when the damper is coupled to a chamber (section \ref{subsec:anmod_results}). These two problems are very different as illustrated in Fig. \ref{fig:bestmdot_RvsGRR}: Problem A consists in maximizing acoustic absorption per unit area, as was done in \cite{hughes1990jfm,putnam1971book,scarpato2013jsv}. When identical dampers are distributed over a surface, the ideal purge mass flow per damper for best normal-incidence absorption depends on the number of dampers per unit area. When this purge mass flow is set, the acoustic resistance of the surface matches the characteristic impedance of the medium. In Fig. \ref{fig:bestmdot_RvsGRR}a and \ref{fig:bestmdot_RvsGRR}c, which give the absolute value of the normal incidence reflection coefficient as in Fig. \ref{fig:R_coeff}, one can see that the optimum mass flow that leads to anechoic condition, is not the same whether 4 or 9 geometrically identical dampers are distributed over the same surface.\\
Problem B deals with the stabilization of an unstable mode in a chamber using damped resonators. The ideal mass flow for achieving the best stability margin does not depend on the density of dampers over the surface enclosing the cavity, but on the volume ratio between the cavity and the dampers. In \ref{fig:bestmdot_RvsGRR}b and \ref{fig:bestmdot_RvsGRR}d, the EP obtained by the connection of a tuned HH resonator to a cavity is not obtained at the same mass flow, when the cavity-to-damper volume ratio is changed. Note also that, although the HH dampers considered for this illustration feature the same geometry as the ones used to clarify Problem A, these mass flow differ from the ones giving  anechoic condition in \ref{fig:bestmdot_RvsGRR}a and \ref{fig:bestmdot_RvsGRR}c. This is due to the fact that the optimization problem is not the same in the two cases: in the first one, one tries to minimize the reflection coefficient of a surface; in the second one, one tries to minimize the real part of the eigenvalues of the coupled system ``chamber-damper''.\\

\noindent In this paper, the damping properties of HH and QW resonators have been investigated. A new linear model for the QW damper impedance has been derived and validated using reflection coefficient measurements. A coupled chamber-damper experiment was set up in order to measure the stability limits of the coupled system for both types of dampers. The damping capabilities of these  dampers have been compared theoretically and experimentally: for comparable volume, the QW damper requires a higher mass flow both for minimizing the reflection coefficient (for which it provides damping in a narrower frequency band as the HH) and for optimizing the stabilization of an acoustic eigenmode. The HH damper is more prone to detuning but also provides better stabilization at very low purge mass flows than a QW damper featuring the same volume. The experiments also allowed the validation of the analytical model describing the coupled system. It was demonstrated that the best damping is achieved at the exceptional point of the coupled system, obtained for tuned dampers and for a critical mass flow whose expression is given as function of the key parameters of the system.

\appendix

\section{Derivation of the analytical model}
\label{sec:appA}
 The problem of coupled cavities has been the topic of numerous studies, which are based on the model given in chapter 10.4 of the book of Morse and Ingard \cite{morse1968book}. In the present case, one of the cavities (the chamber) is much larger than the others (the dissipative resonators). One can for instance refer to the work of Fahy and Schofield \cite{fahy1980jsv}, who derived a model to predict the increase of modal damping induced by a  single damper. Cummings \cite{cummings1992jsv} and Li and Cheng \cite{li2007jsv} adapted the model to a dissipative resonator array. Doria \cite{doria1995jsv} studied the effect of the damper on the mode shape and the influence of different volume ratios between chamber and damper. Subsequent studies extended the use of the model to cavities exhibiting linearly unstable thermoacoustic modes \cite{bellucci2009phd,noiray2012jsv}. Following  the same approach, the pressure in the chamber is expressed as a Galerkin expansion using the orthonormal basis $\boldsymbol{\psi}$ composed of the natural acoustic eigenmodes:

\begin{equation}
p(t, \mathbf{x})=\sum_{i=1}^{\infty} \eta_i (t) \psi_i (\mathbf{x}),
\label{eq:p_enclosure}
\end{equation}

\noindent with $\psi_i (\mathbf{x})$ the natural eigenmodes and $\eta_i (t)$ their amplitude. Assuming that the chamber is equipped with dampers and that, under the effect of a field dependent volumetric source, the pressure field is dominated by one of these modes, the contribution from the other modes can be neglected and one can express the amplitude of that mode in the frequency domain as \cite{noiray2012jsv}:

\begin{normalsize}
\begin{equation}
\begin{split}
\hat{\eta} (s) =  & \frac{s \rho_c c_c^2}{s^2 + \omega_0^2} \frac{1}{V \Lambda} \left( \frac{\gamma -1}{\rho_c c_c^2} \int_V (\hat{Q}_C (s) + \hat{Q}_N (s)) \psi^{*}(\mathbf{x}) dV - \int_{S_d} \hat{\eta} (s) \frac{|\psi(\mathbf{x})|^2}{Z_d(\mathbf{x},s)} dS \right. \\
 & \left. - \int_{S-S_d} \hat{\eta} (s) \frac{|\psi(\mathbf{x})|^2}{Z(\mathbf{x},s)} dS \right).
\label{eq:mod_ampl_begin}
\end{split}
\end{equation}
\end{normalsize}

\noindent In this formula, $\rho_c$ is the air density in the chamber, $c_c$ the speed of sound in the chamber, $\gamma$ the heat capacity ratio, $V$ the volume of the chamber, $\omega_0$ the natural angular frequency of the dominant mode $\psi$ and $\Lambda$ its norm defined as

\begin{equation}
\Lambda = \frac{1}{V} \int_V  |\psi|^{2}dV,
\label{eq:lambda_def}
\end{equation}

\noindent $\hat{Q}_C$ is the \textit{coherent} component of the volumetric source, in the sense that it depends on the acoustic field and therefore on $\hat{\eta}$, while $\hat{Q}_N$ is the \textit{noisy} component of the volumetric source which does not depend on the acoustic field, and which acts as a broadband acoustic forcing; $S_d$ is the area of the chamber walls which is equipped with dampers; $Z_d(\mathbf{x},s)=\hat{\eta} (s) \psi(\mathbf{x}) / \hat{u}(s)$ is the impedance of the dampers and $Z(\mathbf{x},s)$ is the impedance of the chamber walls.
In combustion chambers, thermoacoustic instabilities result from the constructive interaction between the coherent component of the unsteady heat release rate of the flames $Q_C$, and the acoustic field $\eta$. When the acoustic energy produced by the coherent volumetric source  exceeds the dissipation at the boundaries, the thermoacoustic system is linearly unstable. Considering Eq. \eqref{eq:mod_ampl_begin} in the situation where there are no dampers ($S_d=0$), one can express the 
transfer function which links the modal amplitude to the broadband forcing

\begin{equation}
\mathcal{H}_{\text{wod}} (s) =\frac{\hat{\eta} (s)}{\hat{\mathcal{Q}}_N (s)} = \frac{-2 \nu_\text{wod} s}{s^2 -2 \nu_\text{wod} s + \omega_0^2}, 
\label{eq:Hwod}
\end{equation}

\noindent where the subscript ``wod'' stands for ``without dampers'', and where

\begin{equation}
\hat{\mathcal{Q}}_N (s) = \frac{\gamma -1}{-2 \nu_\text{wod} V \Lambda} \int_V \hat{Q}_N (s) \psi^{*}(\mathbf{x}) dV
\end{equation} 

\noindent is the normalized broadband component of volumetric forcing weighted by the mode shape. $\nu_\text{wod}=\beta-\alpha$ is the linear growth/decay rate of the thermoacoustic system that results from the balance between the linear contribution of the source term $\beta$ (which depends on the gain and delay between coherent component of the volumetric source $Q_C$ and pressure $\hat{\eta}$ \cite{culick2006book}) and the natural linear damping of the mode $\alpha$, which results from the impedance at the boundary (last integral in Eq. \eqref{eq:mod_ampl_begin}). The case where $n$ identical HH dampers are coupled to this chamber is now considered. Assuming that the neck of the dampers is compact with respect to the wavelength $2\pi c_c/\omega_0$, the second integral in Eq. \ref{eq:mod_ampl_begin} can be rewritten as

\begin{equation}
\sum_{k=1}^{n} \dfrac{a \psi^2(\mathbf{x}_k)}{Z_H} \hat{\eta}(s),
\end{equation}

\noindent where $a$ is the cross-section of the neck of the dampers and $\mathbf{x}_k$ is the location of the $k^\text{th}$ damper. With $\Psi_d=\sum_{k=1}^{n} \psi^2(\mathbf{x}_k)$ and with the expression of the HH damper impedance

\begin{equation}
Z_H=\rho_d l \frac{s^2 + 2\alpha_H s + \omega_H^2}{s}.
\end{equation}

\noindent One can rearrange Eq. \ref{eq:mod_ampl_begin} as

\begin{equation}
\hat{\eta} (s) = \mathcal{H}_{\text{wod}} \hat{\mathcal{Q}}_N (s) - \mathcal{H}_{\text{wod}} \frac{\rho_c c_c^2}{\rho_d c_d^2} \underbrace{\frac{A L \Psi_d}{V \Lambda}}_{\varepsilon_H^2} \underbrace{\frac{c_d^2 a}{AL l}}_{\omega_H^2} \frac{1}{-2\nu_\text{wod} 2\alpha_H}  \frac{2\alpha_H s}{s^2 + 2\alpha_H s + \omega_H^2} \hat{\eta} (s),
\end{equation}

\noindent with $2\alpha_H = R_H / \rho_d l$ and $R_H$ the resistive term from Eq. \ref{eq:res_H}. Considering that the pressure drop across the damper neck is small ($\bar{p}_c\approx\bar{p}_d$), one has $\rho_d c_d^2=\gamma \bar{p}_d \approx \gamma \bar{p}_c=\rho_c c_c^2$, and one can write the transfer function which links the modal amplitude to the broadband forcing, when the chamber is equipped ``with dampers'':

\begin{equation}
\mathcal{H}_{\text{wd}} (s) =\frac{\hat{\eta}(s)}{\hat{\mathcal{Q}}_N (s)} = \frac{-2 \nu_\text{wod} s \left( s^2 + 2\alpha_H s + \omega_{H}^2 \right)}{\left( s^2 -2 \nu_\text{wod} s + \omega_0^2 \right) \left( s^2 + 2\alpha_H s + \omega_{H}^2 \right) + s^2 \omega_{H}^2 \varepsilon_H^2}.
\label{eq:transfer_func_H_A}
\end{equation}

\noindent In this expression,  $\varepsilon_H$ is the damping efficiency factor. If $\varepsilon_H=0$, then $\mathcal{H}_{\text{wd}}=\mathcal{H}_{\text{wod}}$. In fact, the damping efficiency factor is a mode-shape weighted dampers-to-chamber volume ratio

\begin{equation}
\varepsilon_H^2 = \frac{V_H}{V} \frac{\Psi_d}{\Lambda}
\end{equation} 

\noindent with $V_H=AL$ the volume of one damper. For instance, if all the $n$ dampers are placed at antinodes where $\psi=1$, then $\Psi_d=n$ and $\varepsilon_H^2 = nV_H/V\Lambda$ where one clearly sees the ratio between overall damping volume $nV_H$ and chamber volume $V$. It shows that large $\varepsilon_H$ are achieved for large damping volume, with dampers at antinodes. In the case of the addition of $n$ identical QW dampers, replacing $a$ by $A$ and $Z_H$ by $Z_Q$ from Eq. \ref{eq:QW_impedance_res} and multiplying numerator and denominator by $L \cdot 4 / \pi^2$ yields the same transfer function as Eq. \eqref{eq:transfer_func_H} with subscripts ``$H$'' replaced by ``$Q$'', with $2\alpha_Q = 2R_Q/\rho L$ ($R_Q$ is the QW damper resistance given at Eq. \eqref{eq:res_Q}) and with

\begin{equation}
\varepsilon_Q^2 = \frac{8}{\pi^2} \left(1+\frac{L_\text{cor}}{L_p}\right) \frac{V_Q}{V}  \frac{\Psi_d}{\Lambda},
\end{equation} 

\noindent where $V_Q=AL_p$. The transfer function of the system without (Eq. \eqref{eq:Hwod}) and with (Eq. \eqref{eq:transfer_func_H}) dampers are represented as block diagrams in Fig. \ref{fig:blockdiagram}.

\section*{References}

\bibliography{literaturejsv}

\begin{thebibliography}{10}
\expandafter\ifx\csname url\endcsname\relax
  \def\url#1{\texttt{#1}}\fi
\expandafter\ifx\csname urlprefix\endcsname\relax\def\urlprefix{URL }\fi

\bibitem{zhao2015pas}
D.~Zhao, X.~Y. Li, {A review of acoustic dampers applied to combustion chambers
  in aerospace industry}, Prog. Aerosp. Sci. 74 (2015) 114--130.

\bibitem{lahiri2017jsv}
C.~Lahiri, F.~Bake, A review of bias flow liners for acoustic damping in gas
  turbine combustors, Journal of Sound and Vibration 400 (2017) 564--605.

\bibitem{crocco1969cs}
L.~Crocco, {Research on combustion instability in liquid propellant rockets},
  Symp. Combust. 12~(1) (1969) 85--99.

\bibitem{harrje1972nasa}
D.~Harrje, F.~Reardon, {Liquid propellant rocket combustion instability}, NASA
  SP-194.

\bibitem{sohn2011ast}
C.~H. Sohn, J.~H. Park, A comparative study on acoustic damping induced by
  half-wave, quarter-wave, and helmholtz resonators, Aerospace Science and
  Technology 15~(8) (2011) 606--614.

\bibitem{keller1974nasa}
R.~B. Keller, {Liquid Rocket Engine Combustion Stabilization Devices}, National
  Aeronautics and Space Administration, 1974.

\bibitem{laudien1995paa}
E.~Laudien, R.~Pongratz, R.~Pierro, D.~Preclik, {Experimental procedures aiding
  the design of acoustic cavities}, Prog. Astronaut. Aeronaut. 169 (1995)
  377--402.

\bibitem{acker1994jpp}
T.~L. Acker, C.~E. Mitchell, {Combustion Zone—Acoustic Cavity Interactions in
  Rocket Combustors}, J. Propuls. Power 10~(2).

\bibitem{oschwald2008jpp}
M.~Oschwald, Z.~Farago, G.~Searby, F.~Cheuret, {Resonance Frequencies and
  Damping of a Combustor Acoustically Coupled to an Absorber}, J. Propuls.
  Power 24~(3) (2008) 524--533.

\bibitem{oschwald2011ppp}
M.~Oschwald, M.~Marpert, {On the acoustics of rocket combustors equipped with
  quarter wave absorbers}, Prog. Propuls. Phys. 2 (2011) 339 -- 350.

\bibitem{cardenas2014jpp}
A.~C{\'{a}}rdenas-Miranda, W.~Polifke, {Combustion Stability Analysis of Rocket
  Engines with Resonators Based on Nyquist Plots}, J. Propuls. Power 30~(4)
  (2014) 962--977.

\bibitem{schulze2015jsr}
M.~Schulze, R.~Kathan, T.~Sattelmayer, {Impact of Absorber Ring Position and
  Cavity Length on Acoustic Damping}, J. Spacecr. Rockets 52 (3)~(3) (2015)
  917--927.

\bibitem{zahn2016asme}
M.~Zahn, M.~Schulze, C.~Hirsch, T.~Sattelmayer, {Impact of Quarter Wave Tube
  Arrangement on Damping of Azimuthal Modes}, Proc. ASME Turbo Expo
  2016~(GT2016-56450) (2016) 1--11.

\bibitem{zahn2017asme}
M.~Zahn, M.~Betz, M.~Wagner, N.~V. Stadlmair, M.~Schulze, C.~Hirsch,
  T.~Sattelmayer, {Impact of Damper Parameters on the Stability Margin of an
  Annular Combustor Test Rig}, Proc. ASME Turbo Expo 2017 (2017) 1--11.

\bibitem{joshi1998asme}
N.~D. Joshi, H.~C. Mongia, G.~Leonard, J.~W. Stegmaier, E.~C. Vickers, {Dry low
  emissions combustor development}, in: ASME 1998 Int. Gas Turbine Aeroengine
  Congr. Exhib., American Society of Mechanical Engineers, 1998, pp.
  V003T06A027----V003T06A027.

\bibitem{mongia2003jpp}
H.~C. Mongia, T.~J. Held, G.~C. Hsiao, R.~P. Pandalai, {Challenges and Progress
  in Controlling Dynamics in Gas Turbine Combustors}, J. Propuls. Power 19~(5)
  (2003) 822--829.

\bibitem{garrison1971propconf}
G.~Garrison, G.~Lewis, {The role of acoustic absorbers in preventing combustion
  instability}, 7th Propuls. Jt. Spec. Conf.

\bibitem{hughes1990jfm}
I.~Hughes, A.~P. Dowling, {The absorption of sound by perforated linings}, J.
  Fluid Mech. 218 (1990) 299--335.

\bibitem{burak2009aiaa}
M.~O. Burak, M.~Billson, L.-E. Eriksson, S.~Baralon, Validation of a time-and
  frequency-domain grazing flow acoustic liner model, AIAA journal 47~(8)
  (2009) 1841--1848.

\bibitem{richards2003jpp}
G.~A. Richards, D.~L. Straub, E.~H. Robey, {Passive Control of Combustion
  Dynamics in Stationary Gas Turbines}, J. Propuls. Power 19~(5) (2003)
  795--810.

\bibitem{schlein1999asme}
B.~C. Schlein, D.~A. Anderson, M.~Beukenberg, K.~D. Mohr, H.~L. Leiner,
  W.~Tr{\"{a}}ptau, {Development History and Field Experiences of the First FT8
  Gas Turbine With Dry Low NOx Combustion System}, Proc. ASME Turbo Expo 1999
  (1999) V002T02A039.

\bibitem{bellucci2004jegtp}
V.~Bellucci, P.~Flohr, C.~O. Paschereit, F.~Magni, {On the Use of Helmholtz
  Resonators for Damping Acoustic Pulsations in Industrial Gas Turbines}, J.
  Eng. Gas Turbines Power 126~(2) (2004) 271.

\bibitem{dupere2005jegtp}
I.~D.~J. Dupère, A.~P. Dowling, {The Use of Helmholtz Resonators in a
  Practical Combustor}, J. Eng. Gas Turbines Power 127~(2) (2005) 268.

\bibitem{bothien2013jegtp}
M.~R. Bothien, N.~Noiray, B.~Schuermans, {A Novel Damping Device for Broadband
  Attenuation of Low-Frequency Combustion Pulsations in Gas Turbines}, J. Eng.
  Gas Turbines Power 136~(4) (2013) 041504.

\bibitem{bothien2013asme}
M.~R. Bothien, D.~A. Penelli, M.~Zajadatz, K.~D{\"{o}}bbeling, {On Key Features
  of the AEV Burner Engine Implementation for Operational Flexibility}, Proc.
  ASME Turbo Expo 2013 (2013) 1--9.

\bibitem{esteve2002jasa}
S.~J. Est{\`{e}}ve, M.~E. Johnson, {Reduction of sound transmission into a
  circular cylindrical shell using distributed vibration absorbers and
  Helmholtz resonators}, J. Acoust. Soc. Am. 112~(6) (2002) 2840--2848.

\bibitem{pietrzko2008ast}
S.~J. Pietrzko, Q.~Mao, {New results in active and passive control of sound
  transmission through double wall structures}, Aerosp. Sci. Technol. 12~(1)
  (2008) 42--53.

\bibitem{yu2008jasa}
G.~Yu, D.~Li, L.~Cheng, {Effect of internal resistance of a Helmholtz resonator
  on acoustic energy reduction in enclosures}, J. Acoust. Soc. Am. 124~(6)
  (2008) 3534--3543.

\bibitem{klaus2014aa}
J.~Klaus, I.~Bork, M.~Graf, G.~P. Ostermeyer, {On the adjustment of Helmholtz
  resonators}, Appl. Acoust. 77 (2014) 37--41.

\bibitem{cossalter1994icec}
V.~Cossalter, A.~Doria, F.~Giusto, {Control of Acoustic Vibrations Inside
  Refrigerator Compressors by Means of Resonators}, Int. Compress. Eng. Conf.

\bibitem{gysling2000asme}
D.~L. Gysling, G.~S. Copeland, D.~C. McCormick, W.~M. Proscia, {Combustion
  system damping augmentation with Helmholtz resonators}, J. Eng. Gas Turbines
  Power 122~(2) (2000) 269--274.

\bibitem{soon2012ksnve}
P.~{Soon Hong}, S.~{Sang Hyun}, {Low-frequency noise reduction in an enclosure
  by using a Helmholtz resonator array}, Trans. Korean Soc. Noise Vib. Eng.
  22~(8) (2012) 756--762.

\bibitem{yu2009jsv}
G.~Yu, L.~Cheng, {Location optimization of a long T-shaped acoustic resonator
  array in noise control of enclosures}, J. Sound Vib. 328~(1-2) (2009) 42--56.

\bibitem{zalluhoglu2017jdsmc}
U.~Zalluhoglu, N.~Olgac, {Analytical and Experimental Study on Passive
  Stabilization of Thermoacoustic Dynamics in a Rijke Tube}, J. Dyn. Syst.
  Meas. Control 140~(2) (2017) 021007.

\bibitem{mensah2017jegtp}
G.~A. Mensah, J.~P. Moeck, {Acoustic Damper Placement and Tuning for Annular
  Combustors: An Adjoint-Based Optimization Study}, J. Eng. Gas Turbines Power
  139~(6) (2017) 061501.

\bibitem{zhong2012jasa}
Z.~Zhong, D.~Zhao, {Time-domain characterization of the acoustic damping of a
  perforated liner with bias flow}, J. Acoust. Soc. Am. 132~(1) (2012)
  271--281.

\bibitem{bothien2015aiaa}
M.~R. Bothien, D.~Wassmer, Impact of density discontinuities on the resonance
  frequency of helmholtz resonators, AIAA journal 53~(4) (2015) 877--887.

\bibitem{yang2017aiaa}
D.~Yang, A.~S. Morgans, Acoustic models for cooled helmholtz resonators, AIAA
  Journal (2017) 1--8.

\bibitem{cosic2012jegtp}
B.~Ćosić, T.~G. Reichel, C.~O. Paschereit, {Acoustic Response of a
  Helmholtz Resonator Exposed to Hot-Gas Penetration and High Amplitude
  Oscillations}, J. Eng. Gas Turbines Power 134~(10) (2012) 101503.

\bibitem{noiray2012jsv}
N.~Noiray, B.~Schuermans, {Theoretical and experimental investigations on
  damper performance for suppression of thermoacoustic oscillations}, J. Sound
  Vib. 331~(12) (2012) 2753--2763.

\bibitem{bourquard2018asme}
C.~Bourquard, N.~Noiray, Stability and limit cycles of a nonlinear damper
  acting on a linearly unstable thermoacoustic mode, in: ASME Turbo Expo 2018:
  Turbomachinery Technical Conference and Exposition, American Society of
  Mechanical Engineers, 2018.

\bibitem{shi2016nc}
C.~Shi, M.~Dubois, Y.~Chen, L.~Cheng, H.~Ramezani, Y.~Wang, X.~Zhang,
  {Accessing the exceptional points of parity-time symmetric acoustics}, Nat.
  Commun. 7 (2016) 1--5.

\bibitem{hoffmann2003zamm}
N.~Hoffmann, L.~Gaul, Effects of damping on mode-coupling instability in
  friction induced oscillations, ZAMM-Journal of Applied Mathematics and
  Mechanics/Zeitschrift f{\"u}r Angewandte Mathematik und Mechanik 83~(8)
  (2003) 524--534.

\bibitem{zhu2014prx}
X.~Zhu, H.~Ramezani, C.~Shi, J.~Zhu, X.~Zhang, {PT -symmetric acoustics}, Phys.
  Rev. X 4~(3).

\bibitem{gao2015n}
T.~Gao, E.~Estrecho, K.~Y. Bliokh, T.~C. Liew, M.~D. Fraser, S.~Brodbeck,
  M.~Kamp, C.~Schneider, S.~H{\"{o}}fling, Y.~Yamamoto, F.~Nori, Y.~S. Kivshar,
  A.~G. Truscott, R.~G. Dall, E.~A. Ostrovskaya, {Observation of non-Hermitian
  degeneracies in a chaotic exciton-polariton billiard}, Nature 526~(7574)
  (2015) 554--558.

\bibitem{brandstetter2014nc}
M.~Brandstetter, M.~Liertzer, C.~Deutsch, P.~Klang, J.~Sch{\"{o}}berl, H.~E.
  T{\"{u}}reci, G.~Strasser, K.~Unterrainer, S.~Rotter, {Reversing the pump
  dependence of a laser at an exceptional point}, Nat. Commun. 5~(May) (2014)
  1--7.

\bibitem{achilleos2017prb}
V.~Achilleos, G.~Theocharis, O.~Richoux, V.~Pagneux, {Non-Hermitian acoustic
  metamaterials: Role of exceptional points in sound absorption}, Phys. Rev. B
  95~(14) (2017) 1--9.

\bibitem{xiong2017jasa}
L.~Xiong, B.~Nennig, Y.~Auregan, W.~Bi, {Sound attenuation optimisation using
  metaporous materials tuned on exceptional points}, J. Acoust. Soc. Am.
  142~(4) (2017) 2288--2297.

\bibitem{ryu2015pre}
J.~W. Ryu, W.~S. Son, D.~U. Hwang, S.~Y. Lee, S.~W. Kim, {Exceptional points in
  coupled dissipative dynamical systems}, Phys. Rev. E - Stat. Nonlinear, Soft
  Matter Phys. 91~(5) (2015) 1--6.

\bibitem{ding2016prx}
K.~Ding, G.~Ma, M.~Xiao, Z.~Q. Zhang, C.~T. Chan, {Emergence, coalescence, and
  topological properties of multiple exceptional points and their experimental
  realization}, Phys. Rev. X 6~(2) (2016) 1--13.

\bibitem{mensah2018jsv}
G.~A. Mensah, L.~Magri, C.~F. Silva, P.~E. Buschmann, J.~P. Moeck, Exceptional
  points in the thermoacoustic spectrum, Journal of Sound and Vibration 433
  (2018) 124--128.

\bibitem{tam2001jsv}
C.~K.~W. Tam, K.~Kurbatskii, K.~K. Ahuja, R.~Gaeta, {a Numerical and
  Experimental Investigation of the Dissipation Mechanisms of Resonant Acoustic
  Liners}, J. Sound Vib. 245~(3) (2001) 545--557.

\bibitem{rienstra2015book}
S.~W. Rienstra, A.~Hirschberg, {An Introduction to Acoustics}, Book 0~(0)
  (2015) 296.

\bibitem{yang2016jsv}
D.~Yang, A.~S. Morgans, A semi-analytical model for the acoustic impedance of
  finite length circular holes with mean flow, Journal of Sound and Vibration
  384 (2016) 294--311.

\bibitem{morse1968book}
P.~M. Morse, K.~U. Ingard, {Theoretical acoustics}, Princeton university press,
  1968.

\bibitem{searby2008jpp}
G.~Searby, M.~Habiballah, A.~Nicole, E.~Laroche, {Prediction of the Efficiency
  of Acoustic Damping Cavities}, J. Propuls. Power 24~(3) (2008) 516--523.

\bibitem{nicoud2007aiaa}
F.~Nicoud, L.~Benoit, C.~Sensiau, T.~Poinsot, Acoustic modes in combustors with
  complex impedances and multidimensional active flames, AIAA journal 45~(2)
  (2007) 426--441.

\bibitem{schuermans2004asme}
B.~Schuermans, V.~Bellucci, F.~Guethe, F.~Meili, P.~Flohr, C.~O. Paschereit, {A
  detailed analysis of thermoacoustic interaction mechanisms in a turbulent
  premixed flame}, ASME Turbo Expo 2004 1 (2004) 539--551.

\bibitem{scarpato2012jsv}
A.~Scarpato, N.~Tran, S.~Ducruix, T.~Schuller, {Modeling the damping properties
  of perforated screens traversed by a bias flow and backed by a cavity at low
  Strouhal number}, J. Sound Vib. 331~(2) (2012) 276--290.

\bibitem{noiray2013ijnlm}
N.~Noiray, B.~Schuermans, Deterministic quantities characterizing noise driven
  hopf bifurcations in gas turbine combustors, International Journal of
  Non-Linear Mechanics 50 (2013) 152--163.

\bibitem{li2007jasa}
D.~Li, L.~Cheng, G.~H. Yu, J.~S. Vipperman, {Noise control in enclosures:
  Modeling and experiments with T-shaped acoustic resonators}, J. Acoust. Soc.
  Am. 122~(5) (2007) 2615.

\bibitem{seyranian2005jpa}
A.~P. Seyranian, O.~N. Kirillov, A.~A. Mailybaev, {Coupling of eigenvalues of
  complex matrices at diabolic and exceptional points}, J. Phys. A. Math. Gen.
  38~(8) (2005) 1723--1740.

\bibitem{rees1922amm}
E.~L. Rees, {Graphical discussion of the roots of a quartic equation}, Am.
  Math. Mon. 29~(2) (1922) 51--55.

\bibitem{seyranian2003book}
A.~P. Seyranian, A.~A. Mailybaev, Multiparameter stability theory with
  mechanical applications, Vol.~13, World Scientific, 2003.

\bibitem{kato2013book}
T.~Kato, Perturbation theory for linear operators, Vol. 132, Springer Science
  \& Business Media, 2013.

\bibitem{putnam1971book}
A.~A. Putnam, {Combustion driven oscillations in industry}, Elsevier Publishing
  Company, 1971.

\bibitem{scarpato2013jsv}
A.~Scarpato, S.~Ducruix, T.~Schuller, Optimal and off-design operations of
  acoustic dampers using perforated plates backed by a cavity, Journal of Sound
  and Vibration 332~(20) (2013) 4856--4875.

\bibitem{fahy1980jsv}
F.~J. Fahy, C.~Schofield, {A note on the interaction between a Helmholtz
  resonator and an acoustic mode of an enclosure}, J. Sound Vib. 72~(3) (1980)
  365--378.

\bibitem{cummings1992jsv}
A.~Cummings, {Effects of a resonator array on the sound field in a cavity}, J.
  Sound Vib. 154~(1) (1992) 25--44.

\bibitem{li2007jsv}
D.~Li, L.~Cheng, {Acoustically coupled model of an enclosure and a Helmholtz
  resonator array}, J. Sound Vib. 305~(1-2) (2007) 272--288.

\bibitem{doria1995jsv}
A.~Doria, {Control of acoustic vibrations of an enclosure by means of multiple
  resonators}, J. Sound Vib. 181~(4) (1995) 673--685.

\bibitem{bellucci2009phd}
V.~Bellucci, {Modeling and control of gas turbine thermoacoustic
  pulsations}~(April).

\bibitem{culick2006book}
F.~Culick, Unsteady motions in combustion chambers for propulsion systems,
  Tech. Rep. AG-AVT-039, RTO AGARDograph (2006).

\end{thebibliography}

\end{document}